\documentclass[twocolumn]{aastex7}

\usepackage{multirow}
\usepackage{booktabs}
\usepackage{soul}

\begin{document}

\title{V570\,Per: Characterization of a Benchmark Eclipsing Binary}

\author[0000-0002-9846-3788]{G\"{o}khan Y\"{u}cel}
\altaffiliation{T\"{U}B{\.{I}}TAK-2218 Fellow}
\affiliation{Department of Astronomy and Space Sciences, Faculty of Science, Istanbul University, 34119, Istanbul, T\"{u}rkiye}\email[show]{gokhannyucel@gmail.com}  

\author[0000-0001-9809-7493]{Neslihan Alan}
\affiliation{Department of History of Science, Fatih Sultan Mehmet Vakif University, 34664, Istanbul, T\"{u}rkiye}
\email{neslihan.alan@gmail.com}

\author[0000-0001-9445-4588]{Timothy Banks}
\affiliation{William Rainey Harper College, 1200 W Algonquin Rd, Illinois 60067, USA}
\affiliation{Nielsen, 675 6th Ave, New York, NY 10011, USA}
\email{tbanks@harpercollege.edu}

\author[0000-0003-2575-9892]{Remziye Canbay}
\affiliation{Department of Astronomy and Space Sciences, Faculty of Science, Istanbul University, 34119, Istanbul, T\"{u}rkiye}
\email{rmzycnby@gmail.com}

\author[0000-0002-3125-9010]{Volkan Bak{\i}\c{s}}
\affiliation{Department of Space Sciences and Technologies, Faculty of Science, Akdeniz University, 07058, Antalya, T\"{u}rkiye}
\email{vbakis@akdeniz.edu.tr}

\author[0000-0003-3510-1509]{Sel\c{c}uk Bilir}
\affiliation{Department of Astronomy and Space Sciences, Faculty of Science, Istanbul University, 34119, Istanbul, T\"{u}rkiye}
\email{sbilir@istanbul.edu.tr}


\begin{abstract}
This study presents a comprehensive analysis of the detached binary system V570\,Per through combined photometric, spectroscopic, and astrometric observations. By disentangling the composite spectra, precise fundamental parameters and detailed chemical abundances were determined for both stars. The primary component has a mass of $1.4569_{-0.0100}^{+0.0094}$ $M_{\odot}$, a radius of $1.543_{-0.009}^{+0.012}$ $R_{\odot}$, an effective temperature of $6556_{-26}^{+64}$ K, and a metallicity of $+0.18_{-0.01}^{+0.03}$ dex, while the secondary has a mass of $1.3579_{-0.0089}^{+0.0094}$ $M_{\odot}$, a radius of $1.377_{-0.015}^{+0.013}$ $R_{\odot}$, an effective temperature of $6468_{-21}^{+32}$ K, and a metallicity of $+0.15_{-0.01}^{+0.02}$ dex. The system is estimated to be $577_{-60}^{+60}$ Myr old with a synchronized orbit. The available O-C data imply a minimum mass of 0.57 $M_{\odot}$ or 0.11 $M_{\odot}$ for the third body, corresponding to eccentric and circular orbits, respectively. \texttt{MESA} evolutionary models indicate that the primary will fill its Roche lobe and begin mass transfer to the secondary in about 2.8 Gyr, whereas the secondary will reach the terminal-age main sequence in approximately 3.2 Gyr. The chemical composition of both stars shows remarkable consistency, confirming their common origin, except for calcium, which is significantly enhanced in the primary. Line measurements support this difference and are therefore interpreted as an intrinsic abundance variation rather than an artifact of the analysis. The overall solar-rich metallicity, combined with a relatively low $\alpha$-element content, links V570\,Per to the Galactic thin-disk population. In addition, Galactic orbit analyses of open clusters in the solar neighborhood have revealed that the V570\,Per system may have originated from the Melotte 25 open cluster.


\end{abstract}

\keywords{\uat{Detached binary stars}{375} --- \uat{High resolution spectroscopy}{2096} --- \uat{Galaxy kinematics}{602}  --- \uat{Chemical abundances}{224}  --- \uat{Metallicity}{1031}}


\makeatletter\def\Hy@Warning#1{}\makeatother
\section{Introduction} 

Double-lined spectroscopic detached eclipsing binaries (SB2 dEBs) are among the most valuable astrophysical systems for deriving accurate and direct measurements of fundamental stellar parameters \citep{Southworth_2004}. Space-based photometry from missions such as {\it Kepler} \citep{Borucki_2010} and Transiting Exoplanet Survey Satellite \citep[{\it TESS},][]{Ricker2015}, combined with stable high-resolution spectroscopy, have led to improved estimates compared to earlier decades. A combined analysis of photometric and spectroscopic data enables precise determination of stellar masses, radii, temperatures, and luminosities --- often to better than 1\% accuracy \citep[e.g.][]{Torres2010, eker2021b, Serenelli2021}. These measurements serve as essential benchmarks for testing and calibrating stellar evolution models across a broad range of masses and compositions \citep[e.g.,][]{Munari2001, Torres2010}. Recent studies have further underscored the significance of such systems in constructing empirical relations and verifying theoretical predictions \citep[e.g.,][]{Malkov2003, malkov2007, Gafeira2012, Eker2014, Eker2015, Eker2018, Eker2020, Eker2021a, Eker2022, Eker2023, Eker2024, Eker2025, Moya2018}. High-quality observations of non-interacting SB2 dEBs composed of intrinsically stable components can provide stringent constraints for single-star evolutionary tracks.  

V570\,Per (HD 19457, HIP 14673; $l=145^{\rm o}\!\!.17840, b=-8^{\rm o}\!\!.18863$) was discovered as an eclipsing binary by the HIgh Precision PARallax COllecting Satellite mission \citep[{\rm Hipparcos,}][]{Perryman1997}. It has a short orbital period of approximately 1.9 days and comprises of two main-sequence stars with spectral classifications of F3 and F5 \citep{Munari2001, Marrese2004}. The system's distance was determined using {\it Hipparcos} \citep{ESA1997} trigonometric measurement techniques to be $117^{+15}_{-13}$ pc. High-resolution Echelle spectroscopy and Johnson $BV$ photometry have revealed that the system is well-detached and dynamically relaxed, with both components exhibiting projected rotational velocities ($V_{\rm rot 1,2}\sin i=40\pm1$ and $36\pm1$ km s$^{-1}$) consistent with co-rotation \citep{Tomasella2008}. The obtained masses and radii of the primary ($1.449 \pm 0.006 M_{\odot}$, $1.523 \pm 0.030 R_{\odot}$) and secondary ($1.350 \pm 0.006 M_{\odot}$, $1.388 \pm 0.019 R_{\odot}$) components firmly position both stars within the main-sequence band  \citep{Tomasella2008}. The effective temperatures of the component stars as $T_{1}=6842\pm25$ K and $T_{2}=6562\pm25$ K, along with the metallicity as [M/H] = $+0.02 \pm 0.03$ dex, were derived from $\chi^2$ fitting of high-S/N spectra. A distance of $123 \pm 2$ pc derived from the orbital parameters aligns closely with the revised {\it Hipparcos} estimate of $123 \pm 11$ pc from \cite{van_Leeuwen2007}. Reddening has been independently determined from interstellar Na-I and K-I lines as $E(B-V)=0.023\pm 0.007$ mag, confirming negligible extinction along the line of sight \citep{Tomasella2008}. Additionally, no signs of chromospheric activity or intrinsic variability were detected, reinforcing its suitability as a reference system for testing stellar physics.

V570\,Per is especially noteworthy because the component masses lie in the transitional range where the internal stellar structure changes from convective to radiative cores. This mass regime is sensitive to the treatment of physical processes such as convective core overshooting and element diffusion, which significantly affect the evolutionary tracks. Tailored models from the Bag of Stellar Tracks and Isochrones \citep[BaSTI,][]{Cordier2007} grid computed for the exact observed masses suggest that mild but non-zero core overshooting is required to match the observed properties, with overshooting efficiencies of $\lambda_{\rm OV}=0.14$ and 0.11 for the primary and secondary components, respectively \citep{Tomasella2008}. The age of the system, computed using these BaSTI evolutionary tracks, was reported as $790 \pm 60$ Myr.

The most recent and comprehensive analysis of the F-type eclipsing binary V570\,Per has been carried out by \citet{Southworth2023}, making it the latest benchmark study of this system in the literature. Utilizing high-precision photometric data from two sectors of the {\it TESS}, combined with previously published radial velocity measurements, the study refines the fundamental stellar parameters of the system with notable accuracy. The derived masses and radii as $1.449 \pm 0.006 M_{\odot}$ and $1.538 \pm 0.035 R_{\odot}$ for the primary, and $1.350 \pm 0.006 M_{\odot}$ and $1.349 \pm 0.032 R_{\odot}$ for the secondary component, demonstrate excellent internal consistency; however, the radius estimates remain limited by uncertainties in the spectroscopic light ratio. Interestingly, a delay of approximately 11 minutes in eclipse timing suggests the possible presence of a third body, which we investigated in \S\ref{third_body}. Additional features such as weak starspot activity and a slight discrepancy with {\it Gaia} DR3 orbital solutions further highlight the complexity of the system. Overall, this study reinforces the utility of V570\,Per as a high-value calibrator in stellar structure and evolution research, particularly in the era of precision space photometry.

Building upon previous studies, particularly the precise photometric and spectroscopic characterisation by \citet{Southworth2023}, we present here a new and comprehensive analysis of the V570\,Per system. We combined 35 high-resolution spectroscopic datasets obtained from the ELODIE and Asiago archives with five years of high-precision photometric observations from multiple sectors of the {\it TESS} mission \citep{Ricker2015}. This approach enabled the determination of the system’s fundamental parameters with high accuracy. For the first time in the literature, we performed a detailed chemical abundance analysis of both components, measuring the abundances of more than 20 elements. Furthermore, by employing the Modules for Experiments in Stellar Astrophysics (\texttt{MESA}) stellar evolution code \citep{Paxton2011}, we derived a precise age estimate for the system. The combination of astrometric and spectroscopic data also allowed us to determine its space velocity components, Galactic orbital parameters, and kinematic population membership. Considering the derived age and the Galactic orbital motion, we further constrained the probable birthplace of V570\,Per within the Galaxy.

\section{Data}
\subsection{Photometric Data}

\begin{figure}[ht!]
    \centering
    \includegraphics[width=1\linewidth]{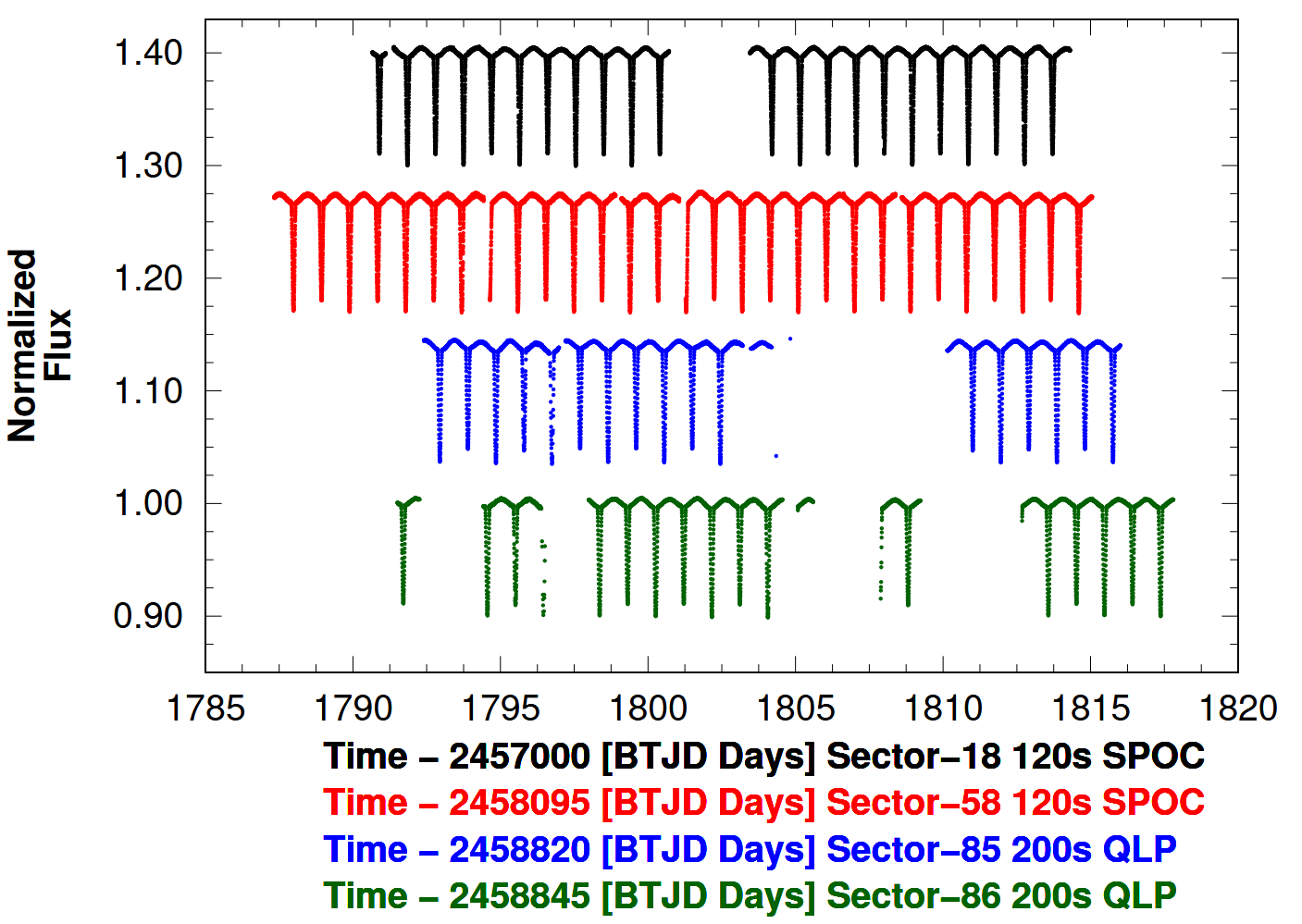}
    \caption{Available photometric \textit{TESS} data for V570\,Per including exposure times for each sector.}
    \label{fig:tess_data}
\end{figure}

There are three different sources for photometric data of V570\,Per in the literature:
\begin{itemize}
    \item \textit{Hipparcos} observed V570\,Per between January 1990 and February 1993. A total of 93 $H$-filter data points are available through CDS services \citep{Perryman1997, Wenger2000}.
    \item Johnson $B$ and $V$ observations were made by \cite{Tomasella2008} using a 28-cm Schmidt-Cassegrain telescope with an Optec SSP5 photometer. A total of 446 and 465 data points, for $B$ and $V\!\!$, respectively, were collected on dates between 2000 and 2003. Details can be found in the relevant paper, while the data are available through CDS\footnote{\url{https://vizier.cds.unistra.fr/viz-bin/VizieR?-source=J/A\%2BA/483/263}}. 
    \item \textit{TESS} observed V570\,Per in four sectors (18, 58, 85, and 86) with exposure times 120s, 120s, 200s, and 200s, respectively. Timing for these sectors is between November 2019 and December 2024. We have downloaded the \textit{TESS} data using \texttt{lightkurve} v2.5.0 with the quality flag setting of ``hard''. It should be noted that sectors 18 and 58 data were selected from the ``SPOC" reduction, and sectors 85 and 86 data were selected from the ``QLP'' reduction. The data used in this study are shown in Figure~\ref{fig:tess_data}.
\end{itemize}

We made use of \textit{TESS} photometry, given its high quality.

\subsection{Spectroscopic Data}

In our spectroscopic analysis, we have used two different datasets from two different spectral sources:

\begin{itemize}
    \item{The first was collected by \cite{Munari2001} using the 1.82-m telescope operated by the Osservatorio Astronomico di Padova atop Mt. Ekar (Asiago). A total of 31 spectra were obtained between December 1998 and December 2000, each covering the range  4550 $< \lambda$ (\AA) $<$ 9000. The dispersion was 0.25 \AA\ per pixel, which with a slit width of 2.0 arcsec, leading to a resolution of 0.42 \AA\  or, equivalently, to a resolving power of $R \sim$ 17000. We obtained the data as the raw spectra and made standard reductions using IRAF \citep{Tody1986, Tody1993}. Unfortunately,  we could not access the data set number 31456 \citep[see Table~2 in][]{Tomasella2008}. However, we have acquired one more observation that was not used in that study.}
    
    \item{The second was retrieved from the ELODIE archive\footnote{\url{http://atlas.obs-hp.fr/elodie/}} \citep{elodie}. The observations were made in January 2001 with the 1.93-m reflector telescope at the Observatoire de Haute-Provence using the ELODIE spectrograph \citep{Baranne}. There are four spectra in the archive, but we have only used three of them since one of the spectra has a low S/N of 27. No additional reduction was required as spectra stored in the ELODIE archive have already been processed.}
\end{itemize}

\section{Data Analyses}

\subsection{Radial Velocity Measurements}

Most (30 out of 34) of the radial velocities (RV) of the components in our spectra have already been measured by \cite{Tomasella2008}, the rest have been measured by following the methods described in \cite{Yucel2025}. Details of the method can be found in the paper. However, to summarize, we have selected seven regions that are mostly unaffected by interstellar extinction lines. Then we built a synthetic spectra grid based on the \texttt{ATLAS9} \citep{Kurucz1979, Castelli2003} and \texttt{SPECTRUM} \citep{Gray1994} codes. Next, we used the same methodology as \citet{Yucel2025}, which simultaneously compares the observed spectrum with the synthetic spectrum, to obtain RVs of the components. The derived RVs, for each region and for each spectrum, are presented in Table~\ref{tab:rv_measured}. The final HJD-corrected values for each component are presented in Table~\ref{tab:rv}.
\begin{table*}
    \setlength{\tabcolsep}{6pt}
    \centering
    \caption{Observed RVs for each component.}
    \label{tab:rv_measured}
    {\footnotesize
    \begin{tabular}{lccccccccc}
    \toprule
            & Julian Date &  \multicolumn{2}{c}{2451894.31796}& \multicolumn{2}{c}{2451927.45764} & \multicolumn{2}{c}{2451930.39762} & \multicolumn{2}{c}{2451931.32282}\\
            \hline
         Reg  &  Wave. Range  &  Primary & Secondary & Primary & Secondary & Primary & Secondary & Primary & Secondary \\
            & (\AA) & \multicolumn{2}{c}{(km s$^{-1}$)} & \multicolumn{2}{c}{(km s$^{-1}$)} & \multicolumn{2}{c}{(km s$^{-1}$)} & \multicolumn{2}{c}{(km s$^{-1}$)} \\
            \hline
        1 & 4810 - 4920 & 47.00$\pm$0.62 & 21.00$\pm$0.60 & 78.32$\pm$1.61 & 6.37$\pm$1.60  & 43.00$\pm$0.10 & 51.00$\pm$0.58 & 57.37$\pm$0.16 & 38.68$\pm$0.91\\
        2 & 4915 - 5025 & 52.63$\pm$0.05 & 20.58$\pm$0.60 & 82.89$\pm$1.00 & 9.42$\pm$1.20  & 43.00$\pm$0.30 & 52.00$\pm$0.79 & 58.42$\pm$0.03 & 36.58$\pm$0.59\\
        3 & 5020 - 5150 & 49.37$\pm$0.16 & 15.26$\pm$1.74 & 81.11$\pm$0.70 & 10.00$\pm$1.11 & 42.00$\pm$0.54 & 52.00$\pm$0.53 & 60.68$\pm$0.26 & 34.47$\pm$0.66\\
        4 & 5145 - 5275 & 44.74$\pm$0.21 & 14.79$\pm$1.02 & 80.79$\pm$0.70 & 9.00$\pm$0.70  & 44.37$\pm$0.95 & 47.42$\pm$0.18 & 56.32$\pm$0.58 & 35.53$\pm$0.66\\
        5 & 5260 - 5405 & 45.26$\pm$0.56 & 17.37$\pm$1.20 & 83.00$\pm$0.09 & 10.00$\pm$0.60 & 43.95$\pm$0.32 & 49.00$\pm$1.03 & 56.32$\pm$0.30 & 36.58$\pm$0.73\\
        6 & 5385 - 5540 & 46.32$\pm$0.34 & 15.26$\pm$0.69 & 82.00$\pm$0.35 & 9.00$\pm$0.80  & 42.89$\pm$0.61 & 50.00$\pm$1.09 & 56.84$\pm$0.55 & 36.58$\pm$0.83\\
        7 & 5525 - 5670 & 50.00$\pm$0.05 & 19.00$\pm$0.54 & 82.00$\pm$0.05 & 11.00$\pm$0.60 & 42.89$\pm$0.91 & 50.00$\pm$0.40 & 57.11$\pm$0.03 & 36.58$\pm$0.60\\
        \hline
        \multicolumn{2}{l}{Weighted Averages} & 49.99$\pm$2.73 & 18.21$\pm$2.62 & 81.92$\pm$0.94 & 9.55$\pm$1.25 & 43.10$\pm$0.58 & 49.51$\pm$1.96 & 57.76$\pm$1.02 & 36.33$\pm$1.21 \\
        \bottomrule
    \end{tabular}
    }
\end{table*}

\begin{table}
    \centering
        \setlength{\tabcolsep}{4pt}
    \renewcommand{\arraystretch}{0.95}
    \caption{Radial velocities: HJD is heliocentric Julian date, RV1 the radial velocities of the primary component, and RV2 the radial velocities of the secondary. One sigma errors are given for both radial velocity columns.} 
    \label{tab:rv}
    {\footnotesize
    \begin{tabular}{lcrrrrr}
    \hline
        HJD & Phase & RV1 & Error & RV2 & Error & S\//N \\
        (day) & ($\phi$) & \multicolumn{4}{c}{(km\,s$^{-1}$)} \\
        \hline
        2451153.44010 & 0.773 & 135.70   & 1.00 & $-98.20$ & 0.60 & 121 \\ 
        2451154.55570 & 0.360 & $-65.80$ & 0.60 & 117.50   & 0.50 & 112 \\ 
        2451155.43590 & 0.823 & 124.80   & 1.00 & $-87.60$ & 0.50 & 126 \\ 
        2451156.46740 & 0.366 & $-63.80$ & 0.50 & 113.90   & 0.40 & 121 \\ 
        2451165.42030 & 0.075 & $-28.40$ & 0.50 &  78.90   & 0.60 & 130 \\ 
        2451166.45460 & 0.619 & 100.70   & 0.50 & $-59.70$ & 0.30 & 143 \\ 
        2451167.42850 & 0.132 & $-61.20$ & 0.80 & 112.40   & 0.50 & 128 \\ 
        2451169.50410 & 0.224 & $-89.80$ & 0.60 & 142.60   & 1.20 & 130 \\ 
        2451197.45530 & 0.928 &  74.30   & 0.30 & $-29.60$ & 0.80 & 137 \\ 
        2451206.33550 & 0.599 &  89.00   & 0.50 & $-47.50$ & 0.90 & 133 \\ 
        2451209.31440 & 0.166 & $-75.50$ & 0.90 & 129.30   & 1.10 & 105 \\ 
        2451217.43460 & 0.438 & $-20.00$ & 0.70 &  69.60   & 0.50 & 110 \\ 
        2451225.36910 & 0.612 &  97.00   & 0.80 & $-55.70$ & 0.40 & 117 \\ 
        2451480.55590 & 0.854 & 114.40   & 0.60 & $-74.20$ & 0.80 & 110 \\ 
        2451485.43390 & 0.420 & $-30.80$ & 0.20 &  82.00   & 0.50 & 102 \\ 
        2451505.46200 & 0.956 &  52.90   & 0.70 & $-10.30$ & 0.90 & 113 \\ 
        2451507.26370 & 0.904 &  87.30   & 0.50 & $-45.80$ & 0.50 & 123 \\ 
        2451507.35520 & 0.952 &  56.10   & 0.40 & $-12.70$ & 0.60 & 137 \\ 
        2451533.49320 & 0.702 & 132.20   & 0.80 & $-93.50$ & 0.50 & 105 \\ 
        2451561.23570 & 0.296 & $-86.30$ & 0.80 & 139.30   & 1.40 &  90 \\ 
        2451561.31710 & 0.339 & $-74.50$ & 0.80 & 125.20   & 1.00 & 137 \\ 
        2451564.22890 & 0.871 & 105.20   & 0.60 & $-66.20$ & 0.80 & 122 \\ 
        2451564.50320 & 0.015 &  12.80   & 0.50 &  35.30   & 1.70 & 111 \\ 
        2451570.26040 & 0.044 &  $-6.50$ & 0.50 &  57.80   & 0.30 & 140 \\ 
        2451894.32236*& 0.515 &  38.76   & 2.62 &  6.98    & 2.73 &  79 \\ 
        2451894.34440 & 0.530 &  44.40   & 0.60 & $-0.80$  & 0.30 &  95 \\ 
        2451895.40910 & 0.090 & $-39.80$ & 0.90 &  88.10   & 0.80 & 113 \\ 
        2451895.50910 & 0.143 & $-66.90$ & 0.70 & 119.00   & 0.70 & 137 \\ 
        2451896.42340 & 0.624 & 102.70   & 0.70 & $-64.20$ & 0.80 & 142 \\ 
        2451896.44570 & 0.636 & 107.80   & 0.70 & $-70.80$ & 0.90 & 139 \\ 
        2451896.46580 & 0.646 & 112.90   & 0.50 & $-75.60$ & 0.90 & 143 \\ 
        2451896.48830 & 0.658 & 116.70   & 1.00 & $-80.90$ & 1.00 & 145 \\ 
        2451927.46031* & 0.947&  59.26   & 0.94 & $-13.11$ & 1.25 &  44 \\ 
        2451930.39994* & 0.494&  19.80   & 0.58 &  26.21   & 1.96 &  55 \\ 
        2451931.32508* & 0.981&  34.34   & 1.02 &  12.91   & 1.21 &  65 \\ \hline
        \multicolumn{7}{l}{*This study}
    \end{tabular}
    }
\end{table}

\subsection{Analysis of RV and LC}
We employed the PHysics Of Eclipsing BinariEs (\texttt{PHOEBE}; \citealt{phoebe1}) v1.0 code to analyze the RV and LC data and derive the parameters of the components of V570\,Per. Both previous studies and an initial inspection of the light curve indicate that no mass transfer occurs between the components; therefore, the system was modeled as a detached binary. During the analysis, the orbital period (\textit{P}) and the effective temperature of the primary component were fixed. The temperature of the primary was set to 6842 K \citep{Tomasella2008}. The following parameters were adjusted: conjunction time ($T_0$), mass ratio ($q$), eccentricity ($e$), argument of periastron ($\omega$), semi-major axis ($a$), systemic velocity ($V_\gamma$), orbital inclination ($i$), effective temperature of the secondary ($T_{\rm 2}$), dimensionless surface potentials of both components ($\Omega_{\rm 1,2}$), and the monochromatic luminosity of the primary ($L_{\rm 1}$). Considering the mass and temperature values for the primary and the secondary components from the previous studies \citep[c.f.,][]{Munari2001, Tomasella2008, Southworth2023}, the gravity-darkening coefficients and bolometric albedos were selected as 0.32 and 0.50 for the primary and the secondary components, respectively \citep{Lucy1967, Rucinski1969}. We adopt the logarithmic limb-darkening law in the analysis and update it for every iteration. During the analysis, we encountered the same issue reported by \citet{Southworth2023}: a significant discrepancy between the conjunction times derived from the RV and LC solutions. This discrepancy suggests the presence of an additional phenomenon affecting the system. \citet{Southworth2023} proposed that this effect might be due to a third component. Consequently, we included third-light contribution as a free parameter in our solutions to test for such an effect; however, no significant contribution was detected that would support the presence of an additional body. To address this challenge, we incorporated the RVs in phase space into our simultaneous solution.

\citet{Southworth2023} noted that, because the eclipses are partial and shallow, they used the spectroscopic light ratio obtained by \citet{Tomasella2008} and adjusted it for the \textit{TESS} band in their analysis. However, the use of spectroscopic light ratios in binary analysis has limitations. Since binary systems are formed within the same molecular cloud, their components are generally expected to share similar chemical abundances. Spectroscopic light ratios are typically derived by comparing the equivalent widths of the same absorption lines, most commonly [Fe/H] lines. While this method generally provides reliable estimates of the component light ratios, it fails when there is a significant difference in chemical abundances between the components (see, e.g.,  \citeauthor{Yucel2025}, \citeyear{Yucel2025}). For this reason, we did not adopt spectroscopic light ratios in our analysis (although see the later \S\ref{sec:determine_abundances}); instead,  we directly adjusted the light contributions of the components. We also detected a very small eccentricity in the system, $e=0.00152$, but our residuals and model fits were inconsistent with this solution. Therefore, to obtain the best-fitting model, we fixed the eccentricity to zero in our final analysis.

\begin{figure*}
  \centering
  \includegraphics[width=0.6\linewidth]{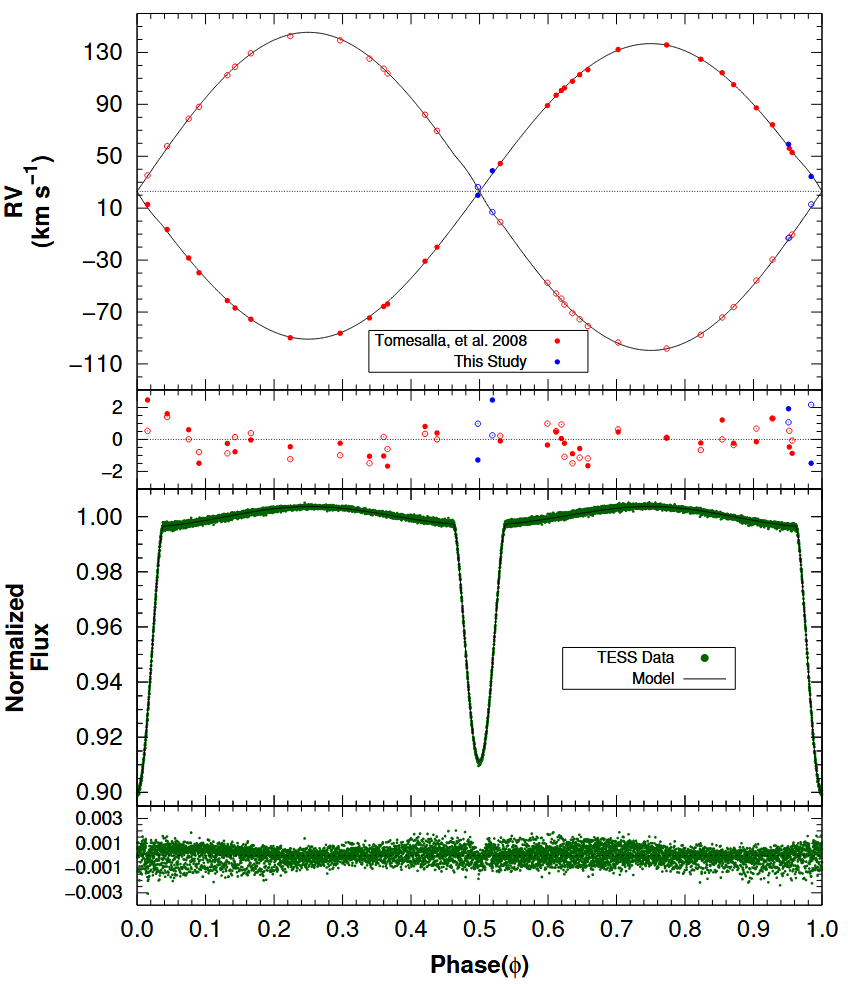}
   \caption{Observed RVs with best-fitting radial velocity curves, and photometric data with LC modelling. Filled and empty circles represent the RVs of the primary and the secondary components of V570\,Per, respectively. In the RV panel, the red and blue dots represent the \citet{Tomasella2008} and ELODIE data, respectively. In the LC panel, the dark green dots represent the \textit{TESS} data, while the black curve is based on the best-fitting LC model for this photometric data set.}
   \label{fig:rvlc}
\end{figure*}

\begin{figure*}
  \centering
  \includegraphics[width=0.9\linewidth]{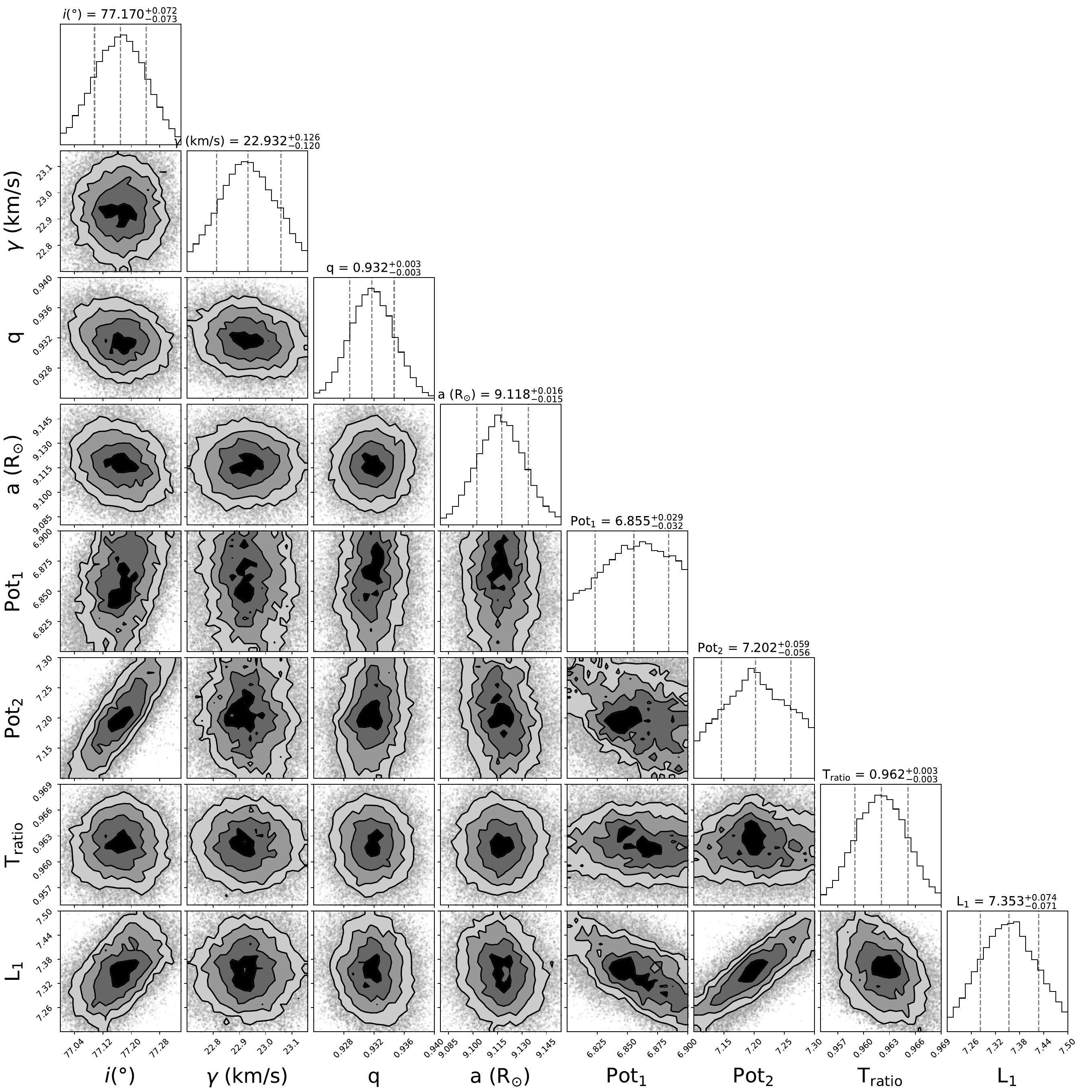}
   \caption{A corner plot of the posteriors for the fundamental parameters of V570\,Per.}
   \label{fig:mcmc}
\end{figure*}

Our final solution and its parameters are presented in Table~\ref{tab:solution}. The RV and LC models are shown in Figure~\ref{fig:rvlc}, and the posterior distribution of our final solution is provided in Figure~\ref{fig:mcmc}.

\begin{table*}
	\setlength{\tabcolsep}{10pt}
	\renewcommand{\arraystretch}{1}
\centering
\caption{Parameters determined for V570\,Per from the analysis of RV and LC data.} 
\label{tab:absolutepar}
\begin{tabular}{lccc}\hline
Parameter                       & Symbol                                & \multicolumn{2}{c}{This Study} \\ 
& & Primary                       & Secondary  \\
\hline
Separation ($R_\odot$)          & $a$                                   & \multicolumn{2}{c}{$9.118^{+0.016}_{-0.015}$}               \\
Mass ratio                      & \emph{q}                              & \multicolumn{2}{c}{$0.932^{+0.003}_{-0.003}$}               \\
Systemic velocity (km s$^{-1}$) & \(V_\gamma\)                          & \multicolumn{2}{c}{$22.932^{+0.126}_{-0.120}$}              \\
Eccentricity                    & \emph{e}                              & \multicolumn{2}{c}{0 (fixed)}                               \\
Orbital inclination ($^\circ$)  & \emph{i}                              & \multicolumn{2}{c}{$77.170^{+0.072}_{-0.073}$}              \\
Temperature ratio               & $T_\mathrm{eff,b}/T_\mathrm{eff,a}$   & \multicolumn{2}{c}{$0.962^{+0.003}_{-0.003}$}               \\
Mass ($M_\odot$)                & \emph{M}                              & $1.4569^{+0.0100}_{-0.0094}$ & $1.3579^{+0.0094}_{-0.0089}$ \\
Radius ($R_\odot$)              & \emph{R}                              & $1.543^{+0.012}_{-0.009}$    & $1.377^{+0.013}_{-0.015}$    \\
Surface gravity (cgs)           & $\log g$                              & $4.225^{+0.009}_{-0.009}$    & $4.293^{+0.013}_{-0.011}$    \\
Light ratio (\textit{TESS})     & $l/l_\mathrm{total}$                  & $0.585^{+0.006}_{-0.006}$    & $0.415^{+0.006}_{-0.006}$    \\
\hline
\label{tab:solution}
\end{tabular}
\end{table*}

\subsection{Spectral Disentangling}

\begin{figure}
    \centering
    \includegraphics[width=\linewidth]{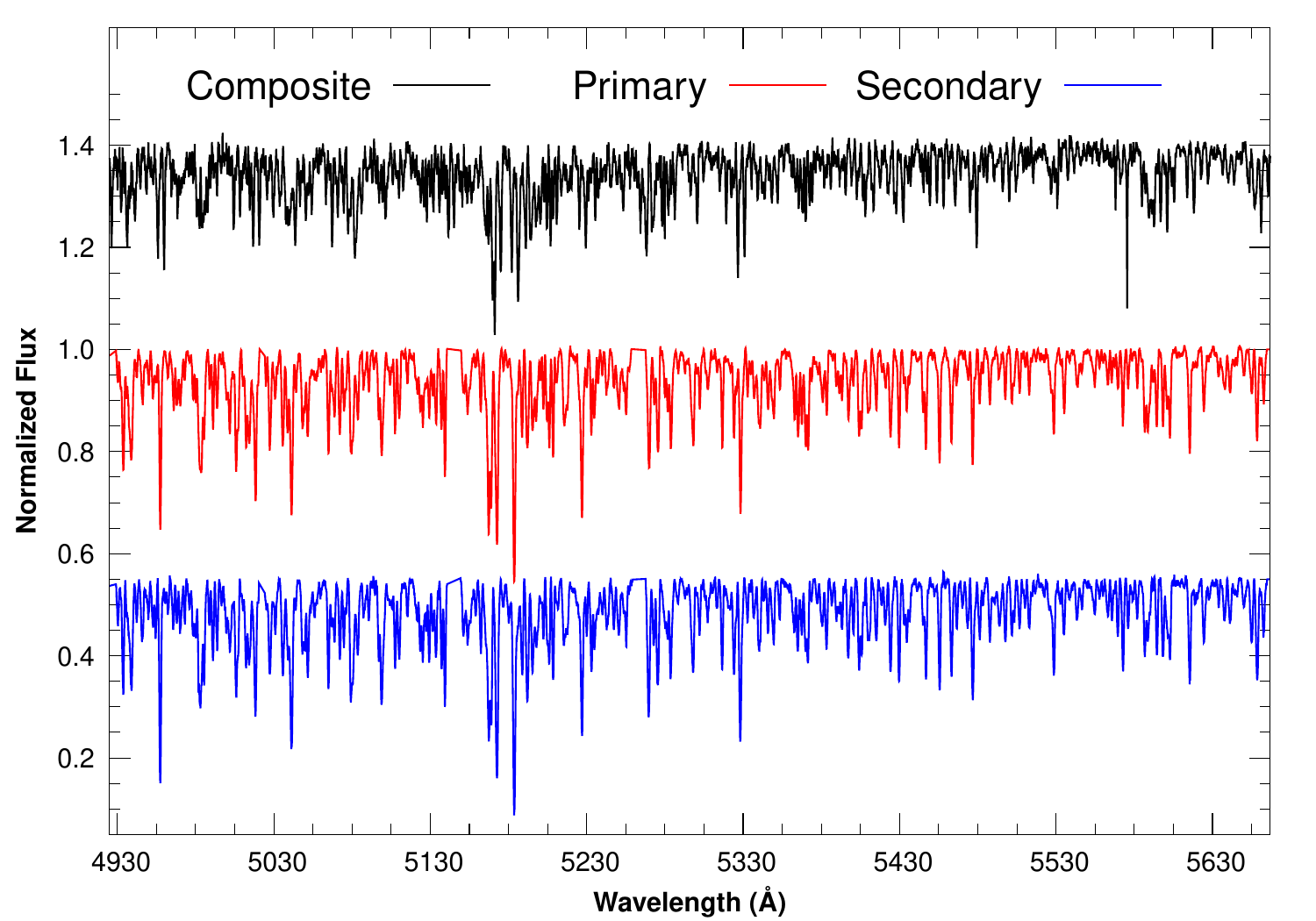}
    \caption{The spectra of the individual components were obtained through spectral disentangling and are shown alongside the observed composite spectrum.}
    \label{fig:disentangle}
\end{figure}

Spectral analysis provides the most reliable method for determining stellar properties such as effective temperature and chemical abundances. However, in binary systems, the observed spectrum is a composite of the contributions from both components, which complicates the analysis. One approach is to construct a grid of synthetic spectra based on stellar atmosphere models, combine them, and compare with the observed spectra to infer the temperatures and abundances of each component \citep{Tomasella2008, Yucel2022}. Nevertheless, the most accurate method for deriving these parameters (one of the primary objectives of this study) is to disentangle the spectra of the individual components from the composite spectrum. For V570\,Per, we utilized two independent spectroscopic datasets. To ensure consistency, we adopted the same procedure as in \citet{Yucel2025}, and applied identical step sizes to the spectra. We then employed the Fourier disentangling code \texttt{FD3} \citep{Ilijic2001} to extract the individual spectra of the components. The resulting disentangled spectra are presented in Figure~\ref{fig:disentangle}.

\subsection{Determination of Temperatures and Chemical Abundances}
\label{sec:determine_abundances}

\begin{figure}
    \centering
    \includegraphics[width=0.90\linewidth]{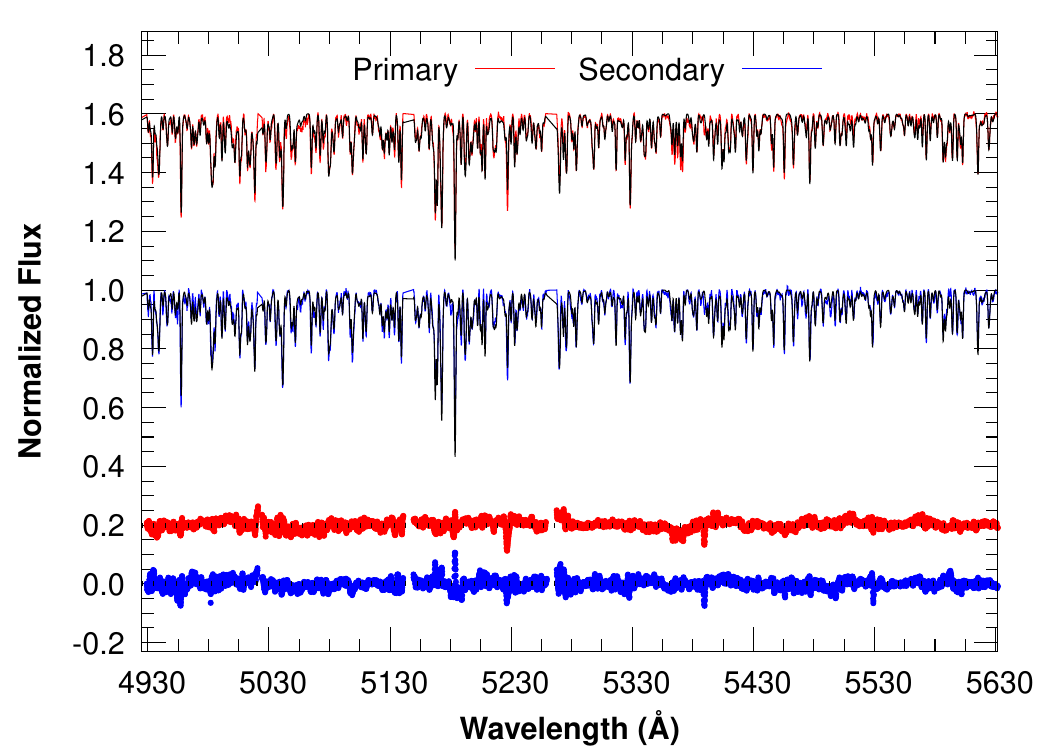}
    \includegraphics[width=0.90\linewidth]{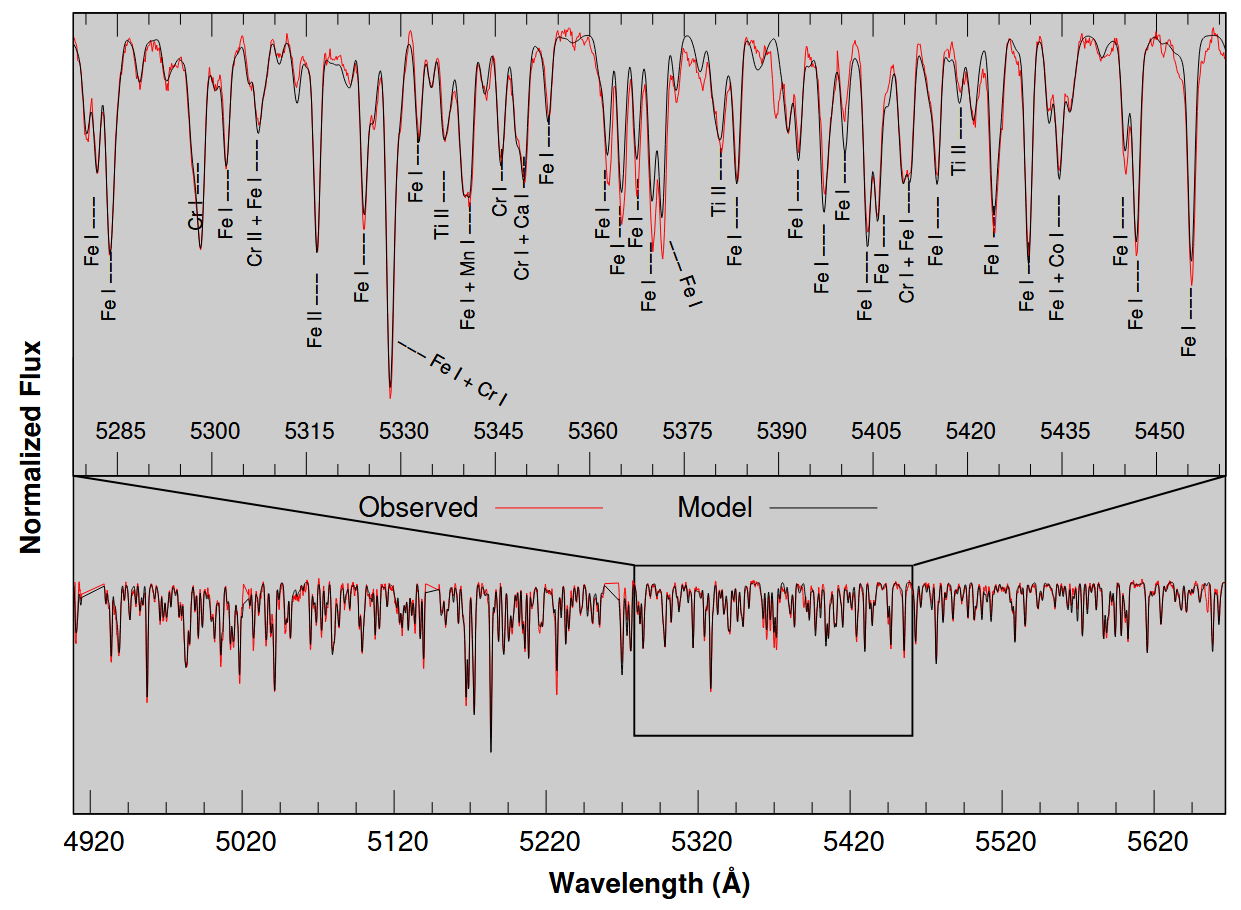}
    \caption{Upper panel shows the disentangled spectra of both components with SP\_Ace models and their residuals, while the lower panel presents a selected region from the primary component, highlighting the spectral lines used for chemical abundance analysis.}
    \label{fig:chem_comparison}
\end{figure}

\begin{figure}
    \centering
    \includegraphics[width=\linewidth]{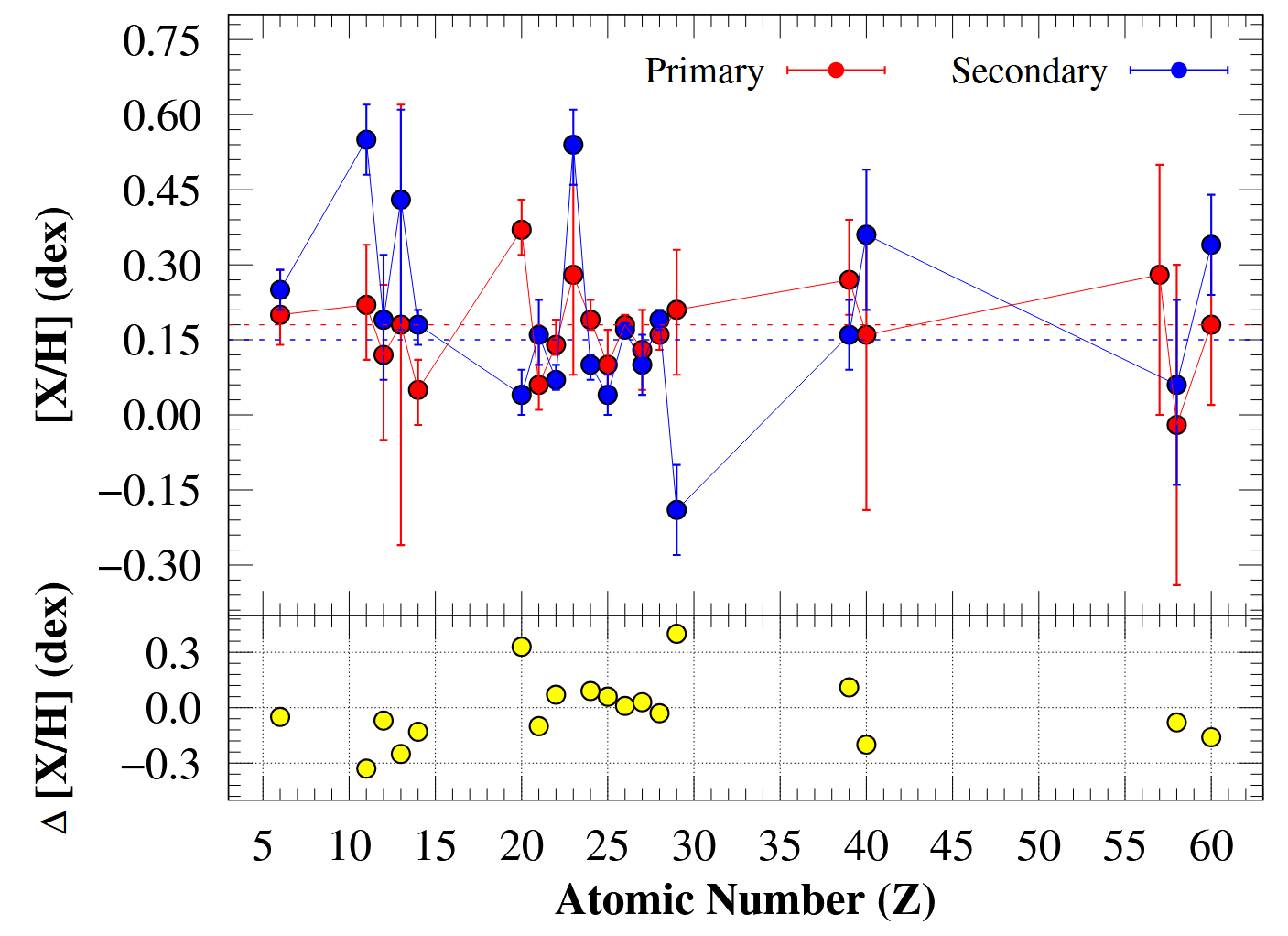}
    \caption{Upper panel: Chemical abundance distributions of the primary and secondary components of V570\,Per as a function of atomic number, shown in red and yellow, respectively. Dashed lines indicate the [Fe/H] value for each component. Lower panel: The differences in chemical abundances between the two components are plotted against atomic number.}
    \label{fig:abundances}
\end{figure}

After successfully disentangling the spectra of the individual components of V570\,Per, we employed \texttt{SP\_Ace} v1.4 \citep{space1, space2} to derive the stellar effective temperatures and chemical abundances. \texttt{SP\_Ace} determines stellar atmospheric parameters ($T_{\rm eff}$, $\log{g}$, and elemental abundances) through spectral fitting. The code simultaneously constructs models based on the General Curve of Growth (GCOG) library, which is generated using \texttt{ATLAS9} and \texttt{SPECTRUM}, and applies them according to the spectral features of absorption lines. Further details can be found in the original references. In our analysis, the $\log{g}$ values for both components were fixed, as they are well constrained by the RV and LC solutions. Our results are summarized in Table~\ref{tab:chem_results}. Figure~\ref{fig:chem_comparison} presents a comparison between the disentangled spectra and the best-fitting \texttt{SP\_Ace} models (upper panel), along with a zoomed-in region of the primary component’s spectrum where individual elements are identified (lower panel). A comparison of the chemical abundances of the two components is shown in Figure~\ref{fig:abundances}. As expected for binary systems formed in the same molecular cloud, the components exhibit no significant overall differences in chemical abundances, except for a few elements such as Na, Al, and Ca. It should be noted that Na and Al abundances are considered to be of lower reliability in \texttt{SP\_Ace} results (see p.\ 12 in \citealt{space2}).

\begin{table}
    \centering
    	\setlength{\tabcolsep}{4pt}
    \caption{SP\_Ace analysis results for each component. The metal abundance ([M/H]) was calculated according to the formula of \cite{Salaris1993}. Letters of ``B" and ``C" as exponents represent the quality of the measurements, and $N$ is the number of lines.}
    \footnotesize
    \begin{tabular}{lcrrrr}
    \hline
        Parameter & Symbol & \multicolumn{2}{c}{Primary} & \multicolumn{2}{c}{Secondary} \\
        \hline
        Temperature & $T_\mathrm{eff}$ (K)& \multicolumn{2}{c}{$6556^{+64}_{-26}$} & \multicolumn{2}{c}{$6468^{+32}_{-21}$} \\
        Signal/Noise & S/N & \multicolumn{2}{c}{982} & \multicolumn{2}{c}{975} \\
        Chi-square & $\chi^2$ & \multicolumn{2}{c}{1.70} & \multicolumn{2}{c}{1.55}  \\
           \hline
\multicolumn{2}{l}{Abundance (dex)} & $N$ & Primary & $N$ & Secondary  \\
        \hline
 \multicolumn{2}{l}{[M/H]}               & 1070 & $0.18^{+0.03}_{-0.01}$  & 1054 & $0.15^{+0.02}_{-0.01}$  \\
 \multicolumn{2}{l}{[Fe/H]}              &  831 & $0.18^{+0.03}_{-0.01}$  &  824 & $0.17^{+0.01}_{-0.01}$  \\
 \multicolumn{2}{l}{[C/H]$^{\rm B}$}     &   15 & $0.20^{+0.09}_{-0.06}$  &   15 & $0.25^{+0.04}_{-0.04}$  \\
 \multicolumn{2}{l}{[Na/H]$^{\rm B}$}    &    4 & $0.22^{+0.12}_{-0.11}$  &    4 & $0.55^{+0.07}_{-0.07}$  \\
 \multicolumn{2}{l}{[Mg/H]}              &    2 & $0.12^{+0.14}_{-0.17}$  &    2 & $0.19^{+0.12}_{-0.12}$  \\
 \multicolumn{2}{l}{[Al/H]$^{\rm B}$}    &    2 & $0.18^{+0.44}_{-0.44}$  &    2 & $0.43^{+0.18}_{-0.26}$  \\
 \multicolumn{2}{l}{[Si/H]}              &   32 & $0.05^{+0.06}_{-0.07}$  &   31 & $0.18^{+0.03}_{-0.04}$  \\
 \multicolumn{2}{l}{[Ca/H]}              &   24 & $0.37^{+0.06}_{-0.05}$  &   24 & $0.04^{+0.05}_{-0.04}$  \\
 \multicolumn{2}{l}{[Sc/H]}              &   26 & $0.06^{+0.09}_{-0.05}$  &   23 & $0.16^{+0.07}_{-0.06}$  \\
 \multicolumn{2}{l}{[Ti/H]}              &  181 & $0.14^{+0.05}_{-0.02}$  &  173 & $0.07^{+0.03}_{-0.02}$  \\
 \multicolumn{2}{l}{[V/H]}               &   28 & $0.28^{+0.18}_{-0.20}$  &   27 & $0.54^{+0.07}_{-0.08}$  \\
 \multicolumn{2}{l}{[Cr/H]}              &  201 & $0.19^{+0.04}_{-0.02}$  &  200 & $0.10^{+0.02}_{-0.03}$  \\
\multicolumn{2}{l}{[Mn/H]}               &   86 & $0.10^{+0.07}_{-0.06}$  &   84 & $0.04^{+0.04}_{-0.04}$  \\
\multicolumn{2}{l}{[Co/H]}               &  127 & $0.13^{+0.08}_{-0.07}$  &  113 & $0.10^{+0.06}_{-0.06}$  \\
\multicolumn{2}{l}{[Ni/H]}               &  140 & $0.16^{+0.03}_{-0.03}$  &  119 & $0.19^{+0.02}_{-0.02}$ \\
\multicolumn{2}{l}{[Cu/H]}               &    6 & $0.21^{+0.12}_{-0.13}$  &    6 & $-0.19^{+0.09}_{-0.09}$ \\
\multicolumn{2}{l}{[Y/H]$^{\rm C}$}      &   18 & $0.27^{+0.12}_{-0.09}$  &   18 & $0.16^{+0.07}_{-0.07}$  \\
\multicolumn{2}{l}{[Zr/H]$^{\rm C}$}     &    9 & $0.16^{+0.21}_{-0.35}$  &    8 & $0.36^{+0.13}_{-0.15}$  \\
\multicolumn{2}{l}{[La/H]$^{\rm B}$}     &   14 & $0.28^{+0.22}_{-0.28}$  &   13 &         ---             \\
\multicolumn{2}{l}{[Ce/H]$^{\rm C}$}     &   44 & $-0.02^{+0.32}_{-0.32}$ &   14 & $0.06^{+0.17}_{-0.20}$  \\
\multicolumn{2}{l}{[Nd/H]$^{\rm C}$}     &   57 & $0.18^{+0.15}_{-0.16}$  &   47 & $0.34^{+0.10}_{-0.10}$  \\
        \hline
            \end{tabular}
    \label{tab:chem_results}
\end{table}

\begin{table*}[ht!]
	\setlength{\tabcolsep}{8pt}
	\renewcommand{\arraystretch}{1.32}
\centering
\caption{Multi-stellar parameters and heuristic errors of V570\,Per.} \label{tab:parameters}
\begin{tabular}{lccc}\hline
Parameter & Symbol  & Primary & Secondary \\
\hline
Equatorial coordinate (Sexagesimal)             & $(\alpha, \delta)_{\rm J2000}$ & \multicolumn{2}{c}{03:09:34.94, +48:37:28.70} \\
Galactic coordinate (Decimal)                   & $(l, b)_{\rm J2000}$  & \multicolumn{2}{c}{145.178402, $-08.188629$} \\
Ephemerides time (d)                            & \emph{T}$_{\rm 0}$    & \multicolumn{2}{c}{$2460650.965777_{-0.00001}^{+0.00001}$} \\ 
Orbital period (d)$^1$                          & \emph{P}              & \multicolumn{2}{c}{$1.90093830_{-0.0000002}^{+0.0000002}$}  \\ 
Separation ($R_\odot$)                          & \emph{a}              & \multicolumn{2}{c}{$9.118^{+0.016}_{-0.015}$} \\ 
Combined visual magnitude (mag)                 & $m_{\rm V}$              & \multicolumn{2}{c}{$8.09_{-0.01}^{+0.01}$}                              \\
Combined visual magnitude$^2$  (mag)            & $m_{\rm TESS}$           & \multicolumn{2}{c}{$7.610_{-0.006}^{+0.006}$}           \\
Combined color index (mag)                      & $B-V$                 & \multicolumn{2}{c}{$0.46_{-0.03}^{+0.03}$}          \\
Color excess (mag)                              & $E(B-V)$              & \multicolumn{2}{c}{$0.01_{-0.01}^{+0.01}$}                \\
Interstellar Extinction (mag)                   & $A_{\rm V}$           & \multicolumn{2}{c}{$0.031_{-0.031}^{+0.031}$}     \\
Interstellar Extinction (mag)                   & $A_{\rm TESS}$        & \multicolumn{2}{c}{$0.021_{-0.019}^{+0.019}$}     \\
Systemic velocity (km\,s$^{-1}$)                & $V_{\gamma}$          & \multicolumn{2}{c}{$+22.932_{-0.120}^{+0.126}$}        \\
Orbital inclination ($^{\circ}$)                & \emph{i}              & \multicolumn{2}{c}{$77.170_{-0.073}^{+0.072}$}      \\
Mass ratio                                      & \emph{q}              & \multicolumn{2}{c}{$0.932_{-0.003}^{+0.003}$}   \\
Eccentricity                                    & \emph{e}              & \multicolumn{2}{c}{$0$ (fixed)}  \\
Argument of perigee (rad)                       & \emph{w}              & \multicolumn{2}{c}{$0$ (fixed)}    \\
Spectral type                                   & Sp                    & F5 V                          & F5.5 V   \\
Metallicity (dex)                               & [M/H]                 & $+0.18_{-0.01}^{+0.03}$       & $+0.15_{-0.01}^{+0.02}$ \\
Mass ($M_\odot$)                                & \emph{M}              & $1.4569_{-0.0100}^{+0.0094}$  & $1.3579_{-0.0089}^{+0.0094}$ \\
Radius ($R_\odot$)                              & \emph{R}              & $1.543_{-0.009}^{+0.012}$     & $1.377_{-0.015}^{+0.013}$ \\
Surface gravity (cgs)                           & $\log g$              & $4.225_{-0.009}^{+0.009}$     & $4.293_{-0.011}^{+0.013}$  \\
Temperature (K)                                 & $T_{\rm eff}$         & $6556_{-26}^{+64}$            & $6468_{-21}^{+32}$   \\
Light ratio (\textit{TESS})                     & $l/l_{\rm{total}}$    & $0.585^{+0.006}_{-0.006}$     & $0.415^{+0.006}_{-0.006}$  \\
Luminosity ($L_\odot$)                          & $\log$ \emph{L}       & $0.5979_{-0.0122}^{+0.0235}$  & $0.4755_{-0.0154}^{+0.0166}$ \\
Bolometric magnitude (mag)                      & $M_{\rm Bol}$         & $3.245_{-0.059}^{+0.031}$     & $3.551_{-0.041}^{+0.039}$   \\
Component's Age (Myr)                           & \emph{t}              & $577_{-64}^{+77}$             & $543_{-126}^{+116}$        \\
System Age (Myr)                                & \emph{t}              & \multicolumn{2}{c}{$577_{-60}^{+60}$}    \\
Individual \emph{TESS} magnitude (mag)          &  $m_{\rm TESS\,1,2}$  & $8.192_{-0.011}^{+0.011}$     & $8.565_{-0.016}^{+0.016}$ \\
Absolute \textit{TESS} magnitude (mag)$^3$      & $M_{\rm TESS\, 1,2}$  & $3.253_{-0.013}^{+0.013}$     & $3.561_{-0.018}^{+0.018}$\\
Bolometric correction (mag)$^3$                 & \emph{BC$_{\rm TESS}$}& $0.204_{-0.111}^{+0.113}$     & $0.406_{-0.114}^{+0.116}$  \\
Bolometric correction (mag)$^3$                 & \emph{BC$_{\rm V}$}   & $0.010_{-0.173}^{+0.176}$     & $0.002_{-0.176}^{+0.173}$   \\
Computed synchronization velocity (km\,s$^{-1}$)& $v_{\rm synch}$       & $41.2_{-0.7}^{+0.7}$          & $36.7_{-1.1}^{+1.1}$   \\
Photometric distance (pc)                       &  \emph{d}             & \multicolumn{2}{c}{$104.6_{-2.5}^{+4.0}$}     \\
{\it Gaia} distance (pc)                        & $d_{\,\varpi}$        & \multicolumn{2}{c}{$120.6_{-0.5}^{+1.0}$}         \\
\hline
\multicolumn{4}{l}{$^1$\cite{Southworth2023}, $^2$\cite{Stassun2019}, $^3$\cite{Eker2023}} \\
\label{table:fund}
\end{tabular}
\end{table*}

\subsection{SED Analysis}

\begin{figure}[!tb]
\centering
\includegraphics[width=0.99\columnwidth]{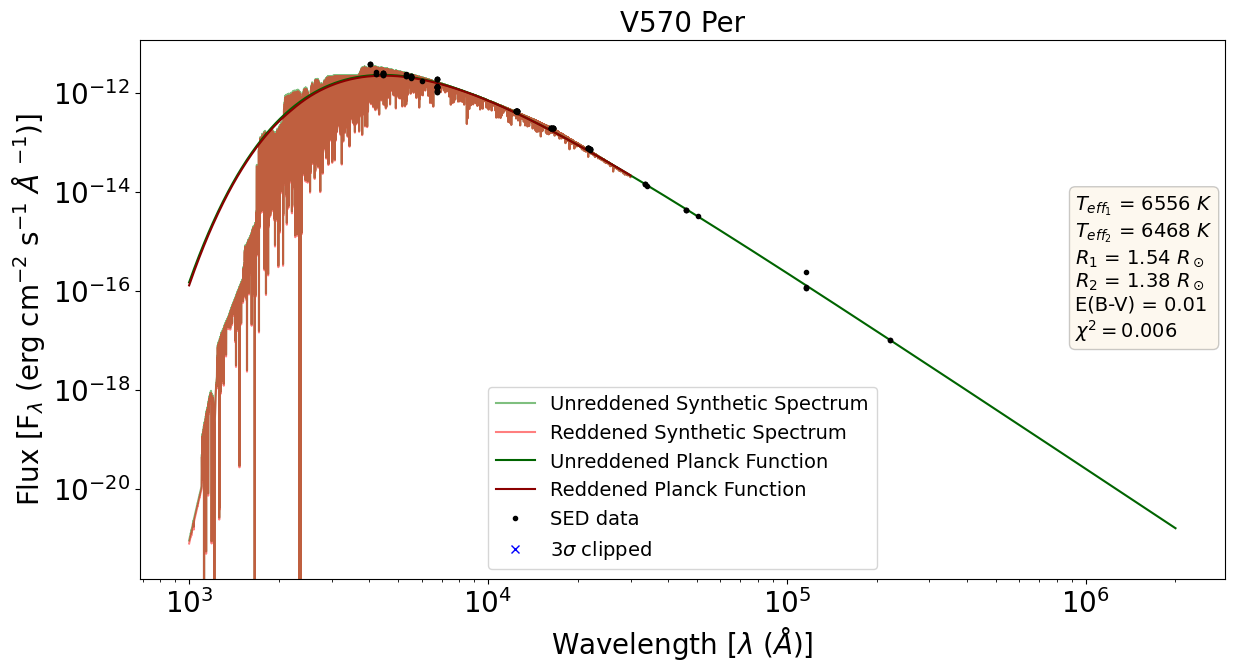}
\caption{SED data (black dots) and the combined synthetic spectra of the components, which are calculated using the absolute parameters of the components and the distance of the system. \label{fig:SED}}
\end{figure}

The $A_{\rm band}$ extinction is estimated from the Spectral Energy Distribution (SED) analysis following the method described in \citet{Bakis2022}, later refined by \citet{Eker2023} for the {\it TESS} pass-band. The best-fitting SED model determines the reddening as $E(B-V)=0.01\pm0.01$ mag, corresponding to $A_{\rm V} =0.031 \pm 0.031$ mag and $A_{\rm TESS} = 0.021 \pm 0.019$ mag. Figure~\ref{fig:SED} presents the SED data alongside synthetic spectra computed using system parameters with \texttt{ATLAS9} models and \texttt{SYNTHE} spectra \citep{Kurucz1993}, demonstrating a strong agreement between the model and observations. In addition, the color excess value of $E(B-V)=0.01\pm0.01$ mag, derived in this study through the SED analysis, is highly consistent with the result of $E(B-V)=0.023 \pm 0.007$ mag obtained by measuring the equivalent widths of the NaI D2 line (5890\AA) in stellar spectra from \citet{Tomasella2008} and applying the calibration of \citet{Munari1997}.

\newpage
\subsection{Third Body}
\label{third_body}

Additional components in eclipsing binary systems cause the system to orbit around the common center of mass formed with the other components. This orbital motion leads to cyclic variations in the observed times of minima due to the so-called light-time effect (LITE). The variation can be represented by the well-known formula given by \citet{Irwin1952}:

\begin{equation}
\Delta t = \frac{a_{12} \sin i_3}{c} \left[ \frac{1 - e_3^2}{1 + e_3 \cos \nu_3} \sin(\nu_3 + \omega_3) + e_3 \sin \omega_3 \right],
\end{equation}
where $a_{12}$ is the semi-major axis of the eclipsing pair around the common center of mass, $i_3$ the inclination of the third-body orbit, $e_3$ the eccentricity, $\omega_3$ the longitude of periastron, $\nu_3$ the true anomaly, and $c$ the speed of light.

For practical purposes, the semi-amplitude $K$ derived from the Eclipse Timing Variation  (ETV) fitting represents the light-travel time corresponding to the projected semi-major axis of the eclipsing pair. Thus, the relation
\begin{equation}
K = \frac{a_{12}\sin i_3}{c}
\end{equation}
holds for circular orbits ($e_3=0$), while for eccentric cases $K$ corresponds to the half the peak-to-peak amplitude of the fitted Irwin model. This relation enables the conversion between the observed timing semi-amplitude and the physical orbital parameters used in the mass-function calculation.

The timing discrepancy of approximately 11 minutes reported by \citet{Southworth2023} for the V570\,Per system may be attributed to such an additional component. Therefore, we collected the available times of minimum light from the literature. The literature data include seven primary and two secondary minima obtained from the Variable Star and Exoplanet Section of the Czech Astronomical Society’s VarAstro web portal\footnote{\url{https://var.astro.cz/en/}}. In addition, we determined 83 new minima (42 primary and 41 secondary) from the {\it TESS} observations.

\begin{figure}[h]
    \centering
    \includegraphics[width=1\linewidth]{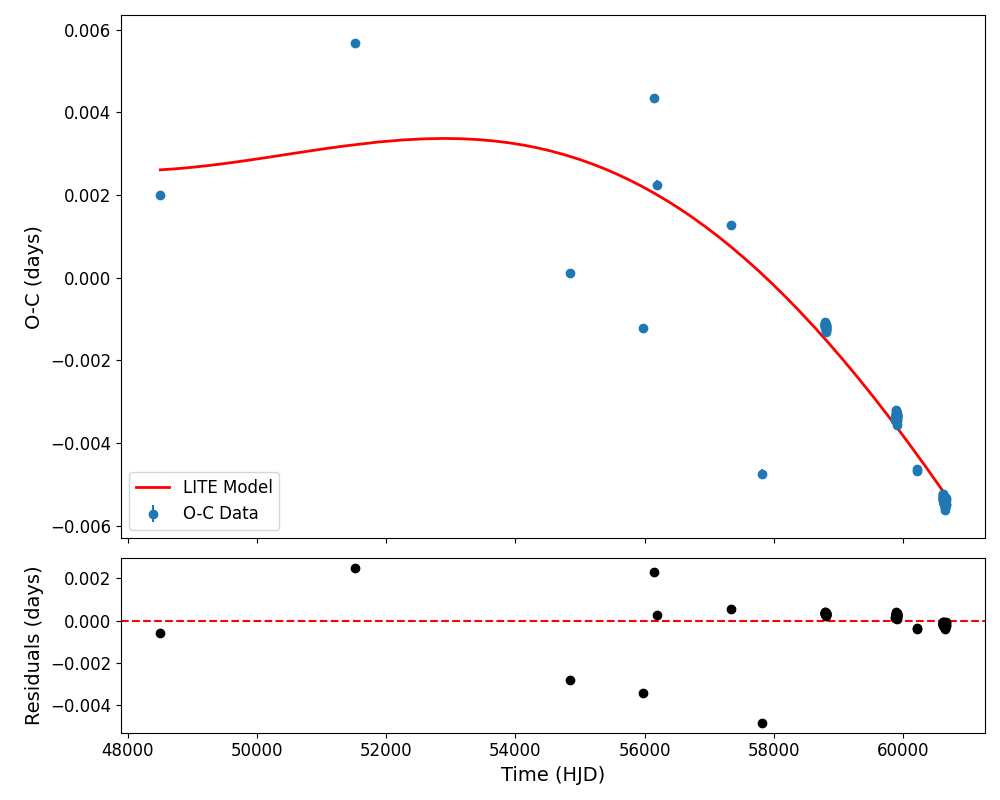}\\
    \includegraphics[width=1\linewidth]{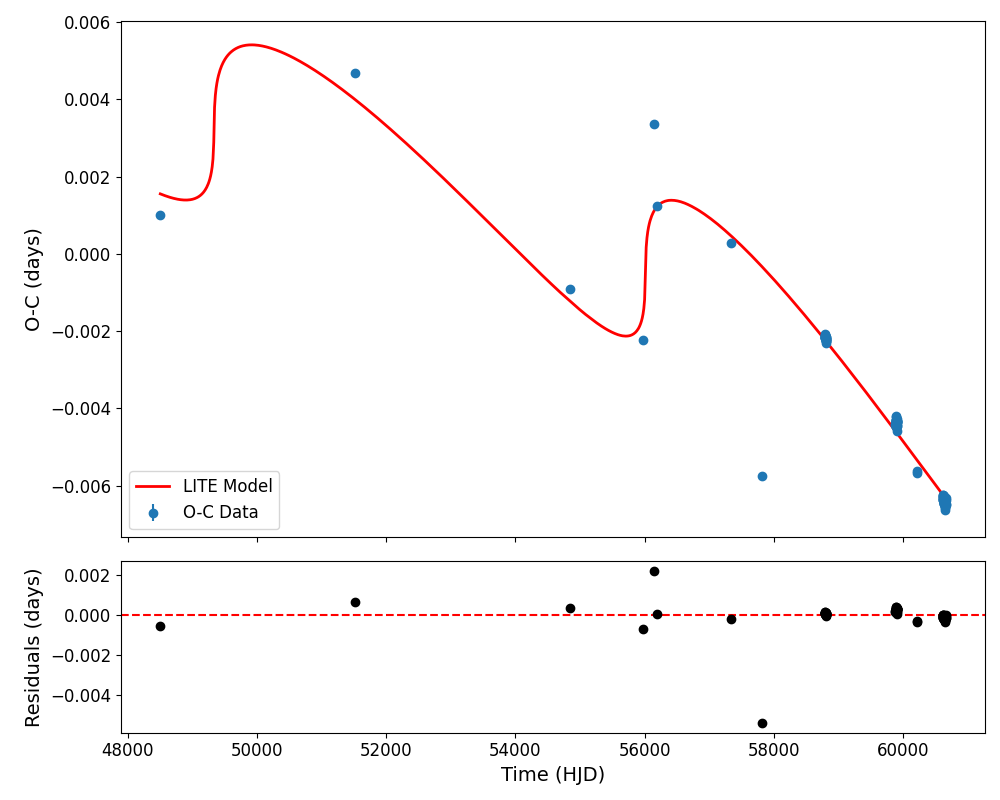}
    \caption{O--C diagrams and two possible orbits, circular (upper panel) and eccentric (lower panel), for the third-body in the system. The residuals from the models are shown at the bottom of each panel.}
    \label{fig:OC_models}
\end{figure}

When the combined times of minima were analyzed using the ETV method, we found that two distinct orbital solutions provided the best representation of the O--C distribution. The circular-orbit model (upper panel of Figure~\ref{fig:OC_models}) yields an orbital period of $P_3 = 25974$~days, a time of periastron passage of $T_0 = 2450437$~(HJD), and a semi-amplitude of $K = 0.0058$~days, corresponding to a mass function of $f(m_3) = 1.97\times10^{-4}\,M_\odot$. This implies a minimum mass of $M_{3,\mathrm{min}} = 0.11\,M_\odot$ for the third body.

The alternative model corresponds to a highly eccentric orbit (lower panel of Figure~\ref{fig:OC_models}). When the eccentricity was left as a free parameter, it reached values as high as $e_3 = 0.98$. In this model, the orbital period is $P_3 = 6680$~days, the time of periastron passage is $T_0 = 2449331$~(HJD), and the semi-amplitude is $K = 0.011$~days. These parameters correspond to a mass function of $f(m_3) = 0.022\,M_\odot$, implying a minimum mass of $M_{3,\mathrm{min}} = 0.57\,M_\odot$ for the third component.

Although both models clearly indicate the presence of an O--C variation, the available data are not sufficient to determine the orbital parameters of the third body conclusively. Future precise timing observations will be crucial for constraining its orbital characteristics.

\section{Evolutionary Status}
\label{sec:evolutionary_status}

We employed the \texttt{MESA}; \citealt{Paxton2011, Paxton2013, Paxton2015, Paxton2018, Paxton2019, Jermyn2023}, version 23.05.1, for the evolutionary calculations. Using its \texttt{binary} module, \texttt{MESA} enables the modeling of the simultaneous evolution of both stars in a binary system, including the calculation of orbital parameters that play a critical role in determining the onset of mass transfer.

\begin{figure*}
    \centering
    \includegraphics[width=0.49\linewidth]{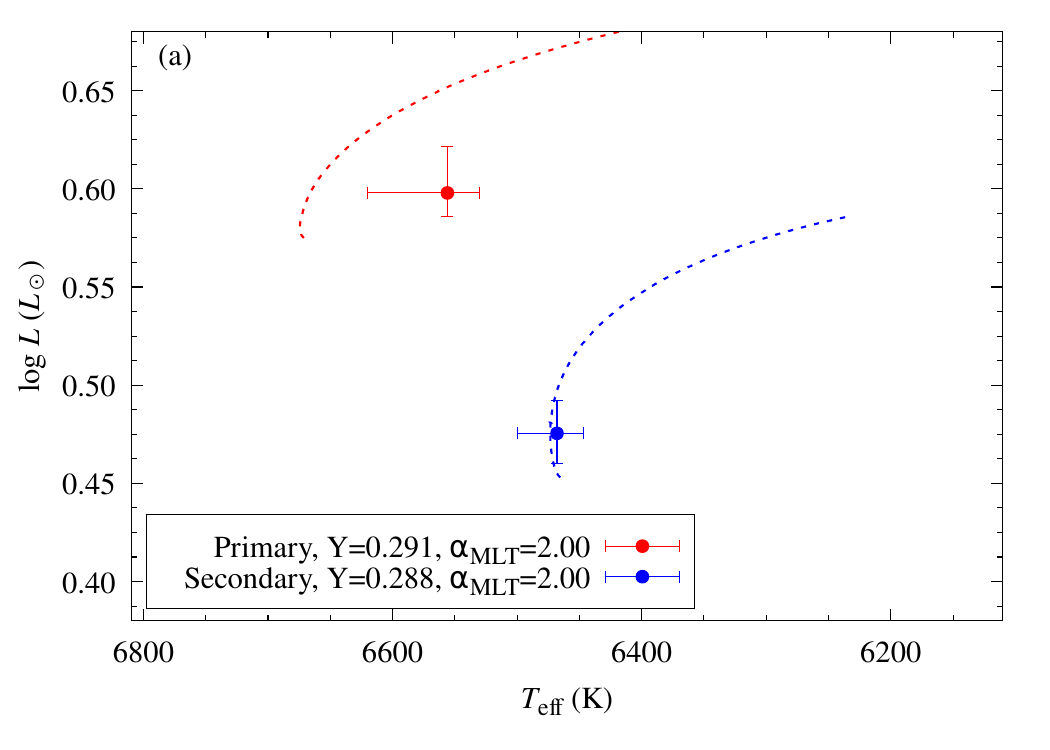}
    \includegraphics[width=0.49\linewidth]{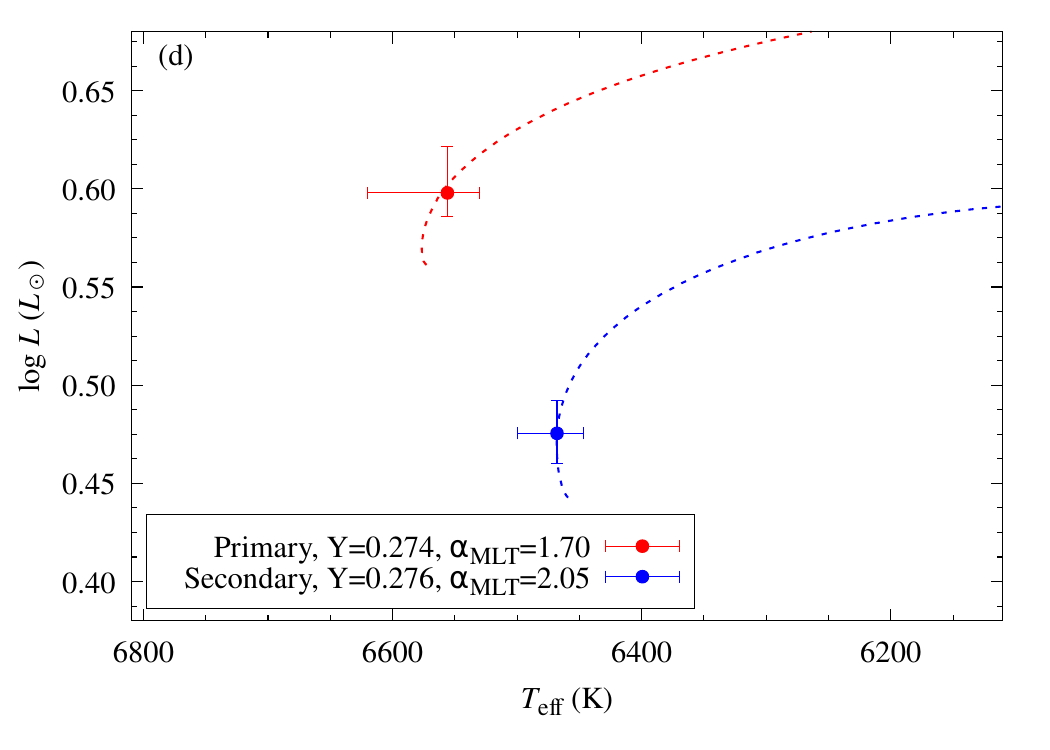}
    \includegraphics[width=0.49\linewidth]{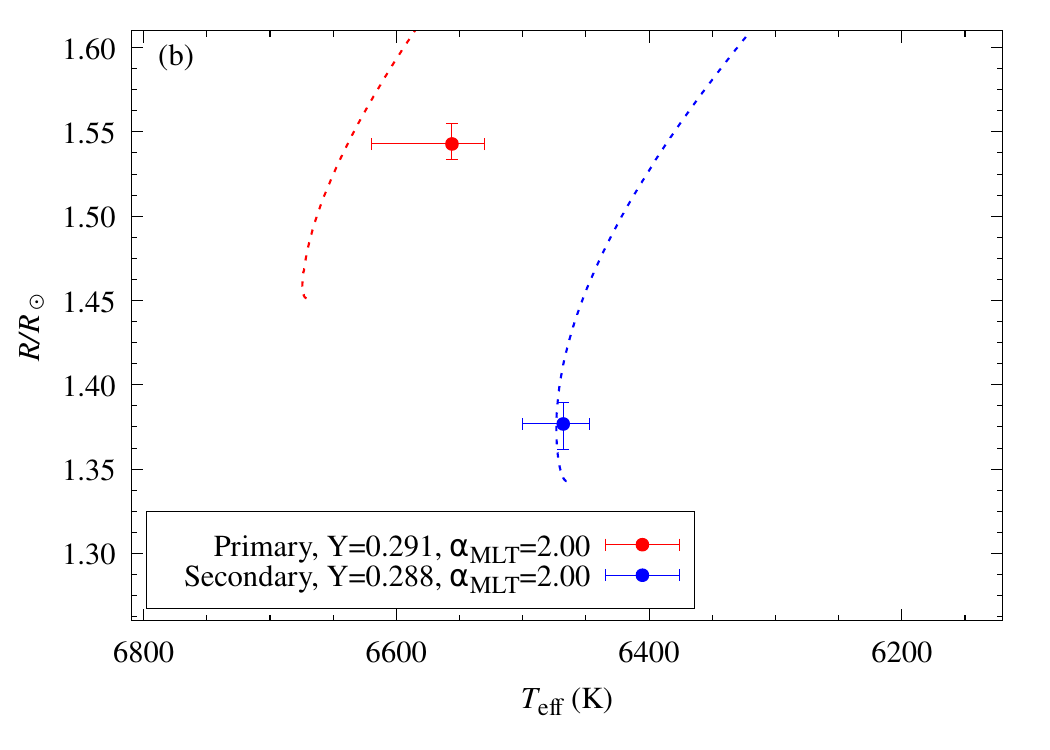}
    \includegraphics[width=0.49\linewidth]{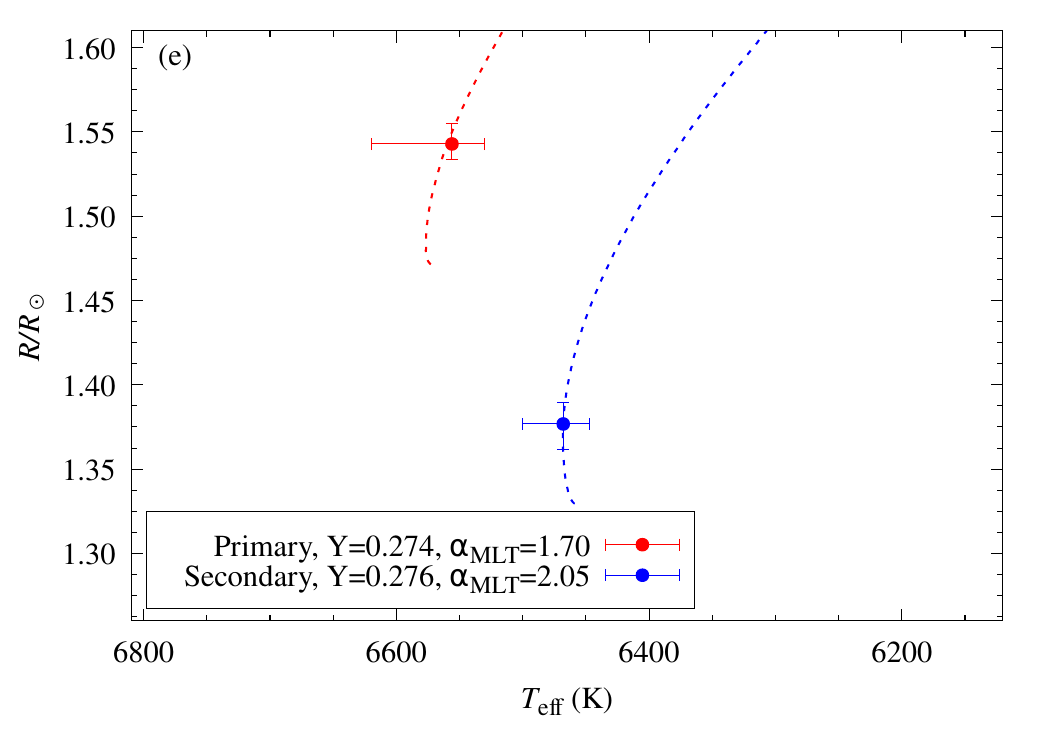}
    \includegraphics[width=0.49\linewidth]{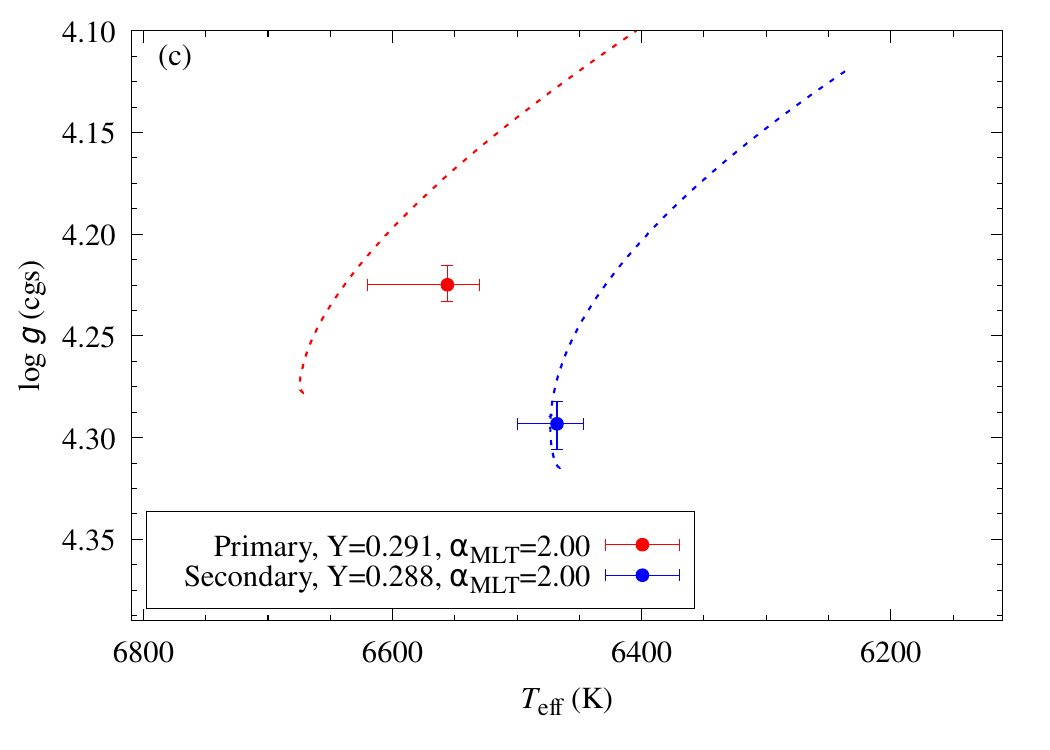}
    \includegraphics[width=0.49\linewidth]{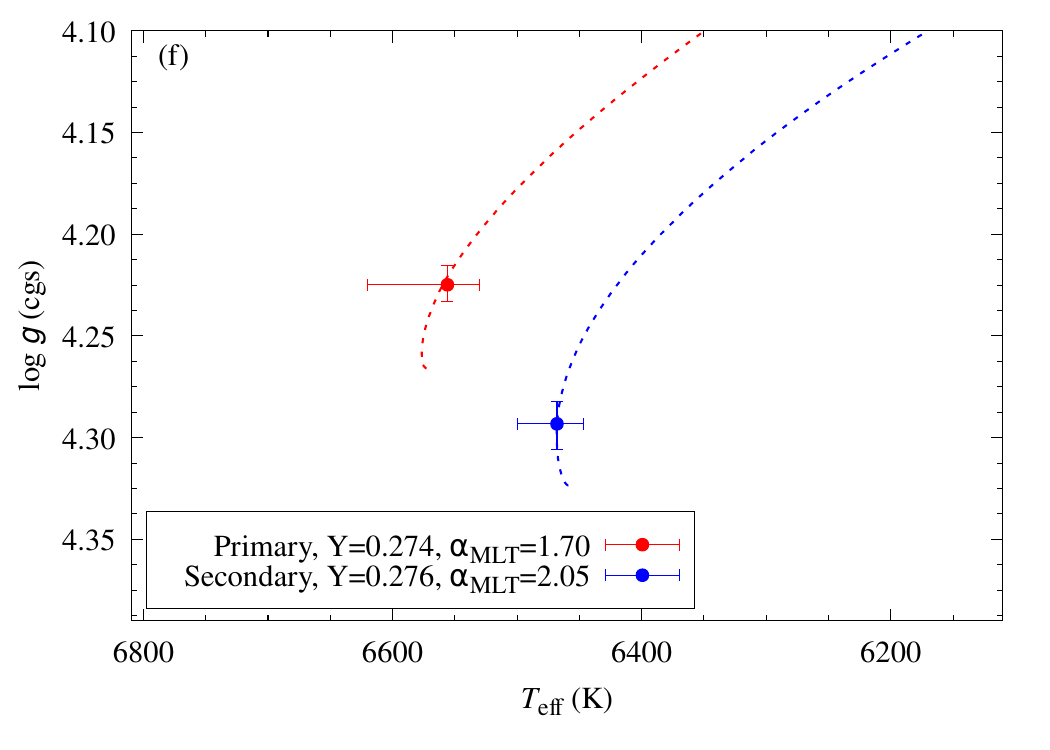}
    \caption{Evolutionary tracks produced in \texttt{MESA} for two different parameter sets: (panels a–c) default values and (panels d–f) adopted values. The evolutionary tracks of the primary and secondary components of V570 Per are shown in the $\log (L/L_{\odot})$–$T_{\rm eff}$ (upper), $R/R_{\odot}$–$T_{\rm eff}$ (middle), and $\log g$–$T_{\rm eff}$ (lower) diagrams, respectively.}
    \label{fig:hr_single}
\end{figure*}


Our initial analysis focused on locating the components of V570\,Per on the HR diagram in order to test the consistency of our results with \texttt{MESA} evolutionary models. Since \texttt{SP\_Ace} adopts the solar composition of \citet{Grevesse1998}, with $Z_{\odot}=0.169$ and $Y_{\odot}=0.2485$, we adopted metallicities of 0.02558 and 0.02387 for the primary and secondary components, corresponding to [M/H] values of 0.18 and 0.15 dex, respectively, in our evolutionary calculations. In \texttt{MESA}, the helium abundance is given by the relation $(Y=0.24+2\times Z)$. Therefore, adopted values for $Y$ are 0.291 and 0.288 for the primary and secondary components, respectively, by default on \texttt{MESA}. Another key parameter in stellar modeling is the mixing length parameter ($\alpha_{\rm MLT})$, which describes convective energy transport. In one-dimensional models, convection is typically treated with $\alpha_{\rm MLT}$. In this framework, $\alpha_{\rm MLT}$ represents the characteristic distance, expressed in units of pressure scale height, that a convective element travels before dissolving \citep[see][]{Joyce2023}. \texttt{MESA} adopts mixing length parameter as 2.0 by default. Our initial evolution models, with using default values, are shown in $\log (L/L_{\odot})\times T_{\rm eff}$ (a), $ R/R_{\odot}\times T_{\rm eff}$ (b), and $\log{g}\times T_{\rm eff}$ (c) diagram, respectively, in Figure~\ref{fig:hr_single}. As can be seen from the diagrams, although the parameters of the second component are consistent with models, the parameters of the primary component are unmatched.

Stellar modeling requires the adoption of appropriately calibrated parameters that describe microscopic and macroscopic mixing processes in stellar interiors. The initial helium abundance ($Y$) is one of the most critical parameters in stellar evolution modeling. Variations in $Y$ strongly influence stellar luminosity, effective temperature, and evolutionary timescales, thereby affecting the positions of stars on the HR diagram. To account for the potential degeneracy between stellar mass and initial helium content \citep[e.g.][]{Fernandes2012, Valle2014, Verma2022}, we treated $Y$ as a free parameter. Following a strategy similar to that adopted by \citet{Lebreton2014} in their single-star modeling, we explored a range starting just below the primordial helium abundance $Y=0.245$ \citep{Peimbert2007} and extending up to $Y=0.300$, with a step size of 0.002. This allowed us to capture the sensitivity of evolutionary tracks to plausible variations in the helium content. Another key parameter in stellar modeling is the mixing length parameter ($\alpha_{\rm MLT})$, which describes convective energy transport. In one-dimensional models, convection is typically treated with $\alpha_{\rm MLT}$. In this framework, $\alpha_{\rm MLT}$ represents the characteristic distance, expressed in units of pressure scale height, that a convective element travels before dissolving \citep[see][]{Fernandes2012, VandenBerg2014, Joyce2023}. Although its exact value is model-dependent and difficult to constrain, $\alpha_{\rm MLT}$ has a major influence on the predicted locations of stars on the HR diagram \citep{Lastennet2003}.

Initially, the default helium values were broadly consistent with our findings; however, within the observational uncertainties, the final adopted values of $Y$ and $\alpha_{\rm MLT}$ were chosen to reproduce the observed positions of the stars in the HR diagram (see Figure~\ref{fig:hr_single} for default values, (a), (b), and (c), and for the adopted values, (d), (e), and (f), respectively). The analysis yielded an age of $577_{-64}^{+77}$ Myr for the primary star ($Y = 0.274$, $\alpha_{\rm MLT}= 1.70$) and $543_{-126}^{+116}$ Myr for the secondary star ($Y = 0.276$, $\alpha_{\rm MLT}=2.05$). Taking the weighted mean of the two components, the system age was determined to be $t = 577\pm 60$ Myr.

The analysis shows that V570\,Per is a detached binary system, with both components still in the main-sequence (MS) phase. This implies that no mass transfer has occurred since the components entered the MS. Consequently, the present-day masses of the stars are essentially the same as their initial values at formation. The only parameters that may have evolved since formation are the orbital elements, besides stars themselves. Given that our RV-LC analysis yielded a zero eccentricity, it is not possible to determine when the orbit of V570\,Per was circularized. Therefore, in our evolutionary calculations, we assumed that the system was formed in a circular orbit.

\begin{figure}
    \centering
    \includegraphics[width=1\linewidth]{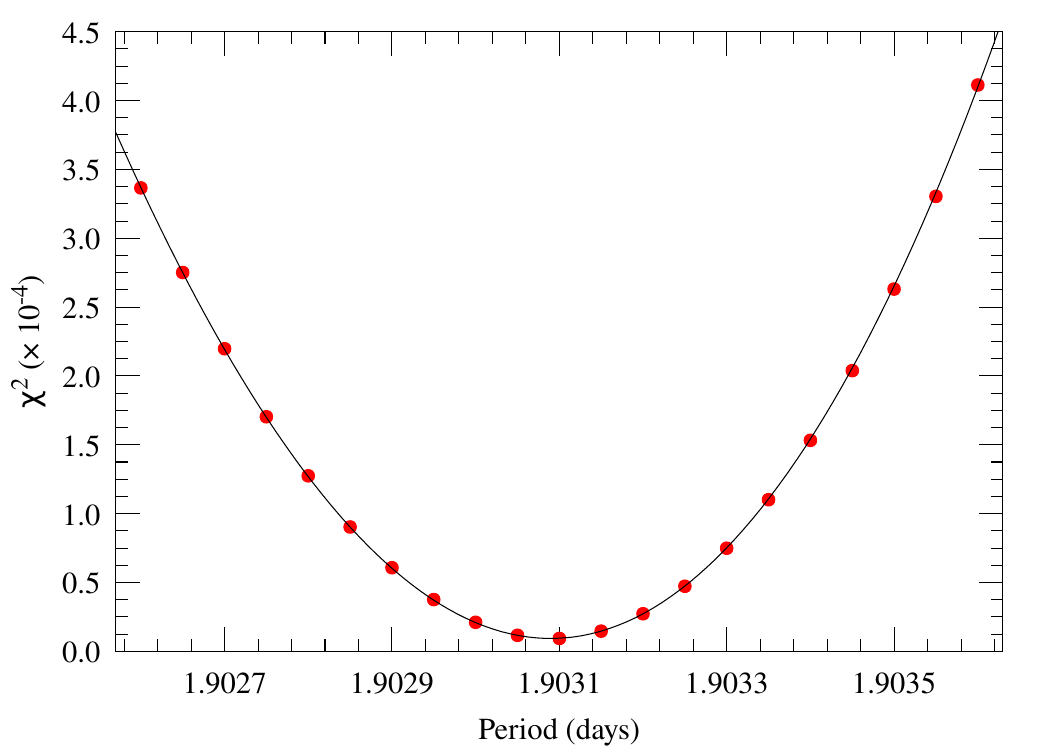}
    \caption{The result of the grid search yielded the best initial orbital condition, with an orbital period of $P=1.90310$ day. Goodness of the fit ($\chi^2$) is plotted against the tested period values, with the minimum $\chi^2$ indicating the best fit period.}
    \label{fig:period}
\end{figure}

\begin{figure}
    \centering
    \includegraphics[width=1\linewidth]{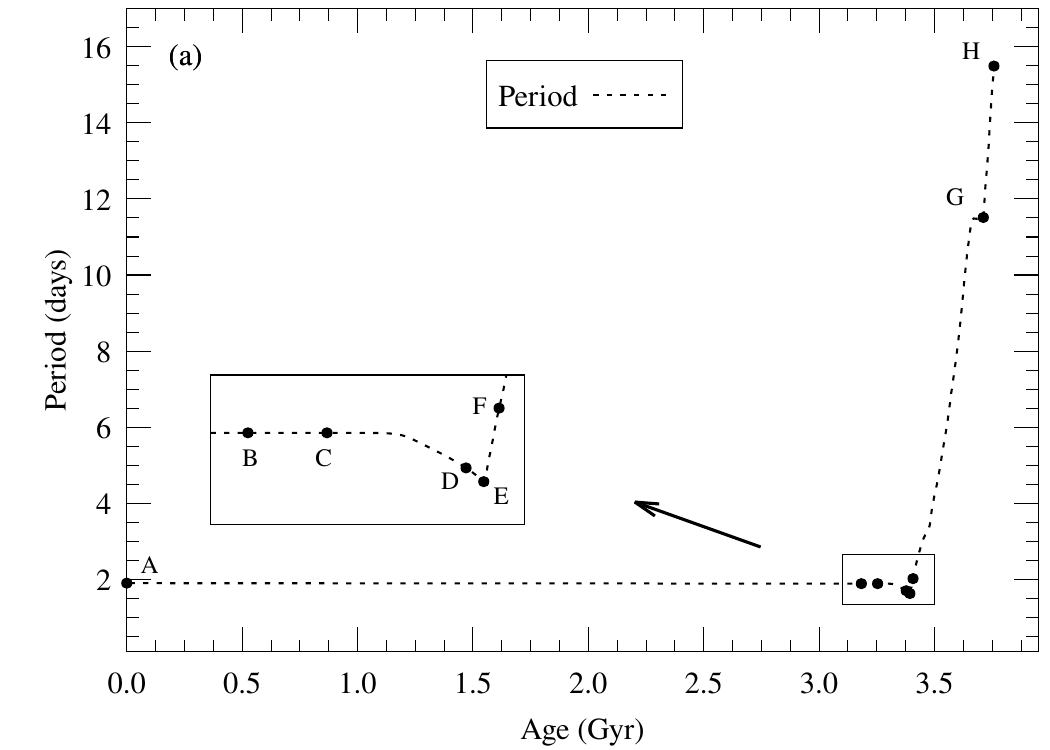}\\
    \vspace{5mm}
    \includegraphics[width=1\linewidth]{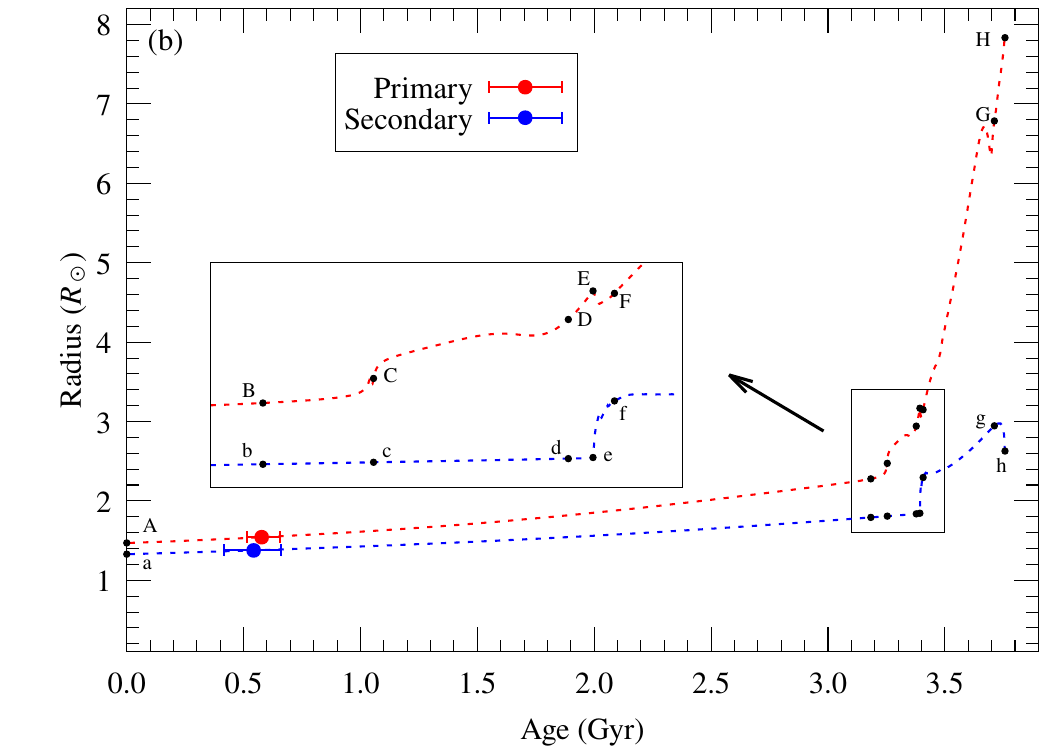}
    \caption{Change of orbital period (a), and radius (b) of the components of V570\,Per with time.}
    \label{fig:radius}
\end{figure}

\begin{figure}
    \centering
    \includegraphics[width=1\linewidth]{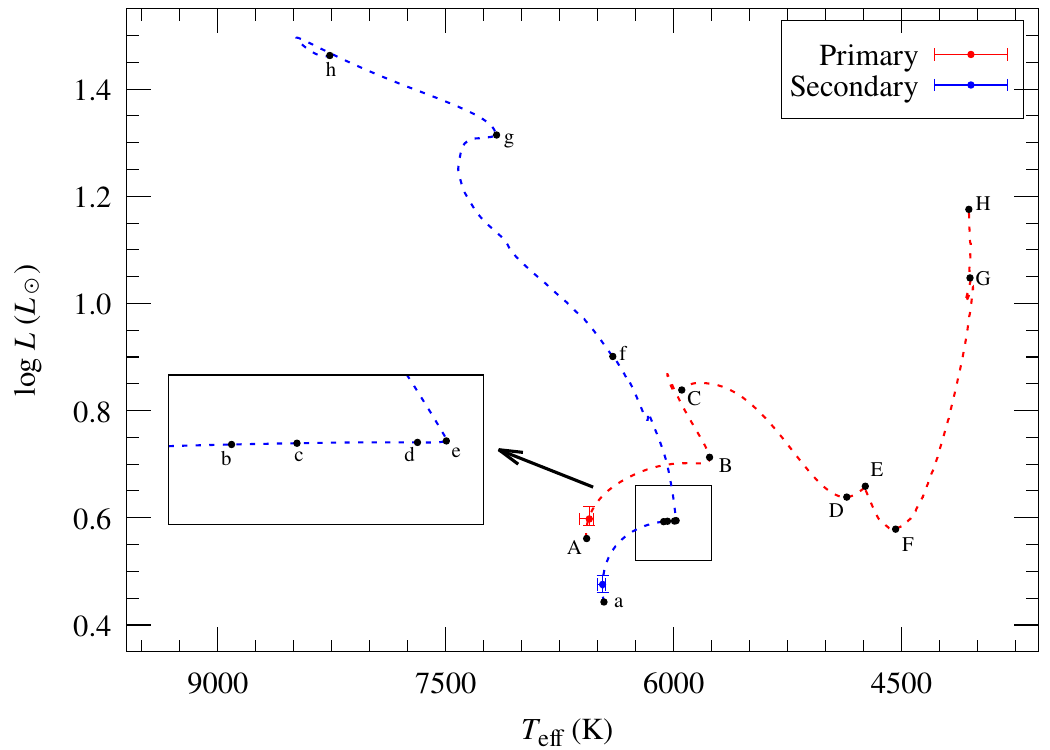}
    \caption{\texttt{MESA} evolutionary tracks of the primary and secondary components of V570\,Per displayed in the $\log (L/L_{\odot})\times T_{\rm eff}$ diagram, showing their positions and evolutionary stages.}
    \label{fig:hr}
\end{figure}

The next step was to determine the initial orbital conditions. To this end, we performed a grid search, a method widely used in the literature \citep[e.g.,][]{Rosales, Soydugan, Soydugan2020, Yucel2022, Yucel2024, Yucel2025}. In this approach, evolutionary calculations were carried out starting from a randomly selected initial orbital period and continued until the present-day period, $1.90093830 \pm 0.00000002$ \citep{Southworth2023}, was reached. A $\chi^2$ analysis was then performed using the current radii and temperatures of the components. The best-fit model corresponded to an initial orbital period of $1.90310$ days (Figure~\ref{fig:period}). In low-mass components, angular momentum loss via magnetic braking can drive orbital decay and synchronization on timescales that depend sensitively on stellar magnetic field strength and wind properties \citep{Skumanich1972, Verbunt1981, Rappaport1983}. In our calculations, we accounted for magnetic braking, given that both components possess convective envelopes \citep{Rappaport1983}. Tidal dissipation acts to synchronize rotation and circularize the orbit on characteristic timescales depending on the stellar structure and separation \citep{Zahn1977}. For tidal synchronization, we adopted the “Orb period” option, which enforces synchronization on the timescale of the orbital period. Roche lobe radii were computed following the prescription of \citet{Eggleton1983}, while mass transfer rates in Roche lobe–overflowing binaries were calculated according to \citet{Kolb1990}. After determining the initial orbital period, we carried out new evolutionary calculations starting from the derived value of 1.90310 days and continued them until the secondary component reached the Terminal-Age Main Sequence (TAMS) phase. The mean mass-transfer efficiency coefficients were adopted following \citet{Paxton2015}, \citet{Rosales}, and \citet{Soydugan}, with values of 0.4, 0.1, and 0.1 for $\alpha$, $\beta$, and $\gamma$, respectively. Finally, our calculations indicate that the system will probably start mass transfer in about 2.82 Gyr. The variations in orbital period and component radii throughout the evolution are shown in Figure~\ref{fig:radius}. A more detailed evolutionary track of both components, including characteristic timescales, is provided in Table~\ref{tab:timestamps} and shown in Figure~\ref{fig:hr}.

\begin{table*}[!ht]
\setlength{\tabcolsep}{5pt}
\centering
    \caption{Detailed evolution of V570\,Per with time-stamps.}
        \begin{tabular}{llcccccccc}
    \toprule
        Mark & \multirow{2}*{Evolutionary Status} & Age & $P$  & \multicolumn{3}{c}{Primary} & \multicolumn{3}{c}{Secondary} \\
       \cline{5-10}
         (Pri/Sec) & & (Myr) & (day) & $T_\mathrm{eff}$ (K) & $\log L$ ($L_\odot$)  & $R$ ($R_\odot$) & $T_\mathrm{eff}$ (K) & $\log L$ ($L_\odot$)  & $R$ ($R_\odot$)\\
        \hline
A/a & ZAMS & 0 & 1.903 & 6573 & 0.561 & 1.472 & 6459 & 0.443 & 1.330 \\ 
        B/b & Primary core contraction & 3183 & 1.891 & 5764 & 0.713 & 2.278 & 6066 & 0.593 & 1.792 \\ 
        C/c & Primary TAMS & 3254 & 1.891 & 5947 & 0.838 & 2.473 & 6041 & 0.593 & 1.808 \\ 
        D/d & Entering RGB & 3378 & 1.703 & 4861 & 0.639 & 2.941 & 5995 & 0.594 & 1.837 \\ 
        E/e & Starting of mass transfer & 3393 & 1.630 & 4739 & 0.659 & 3.600 & 5984 & 0.595 & 1.845 \\ 
        F/f & Re-entering RGB & 3407 & 2.023 & 4539 & 0.579 & 3.149 & 6401 & 0.901 & 2.294 \\ 
        G/g & Secondary core contraction & 3712 & 11.511 & 4050 & 1.048 & 6.785 & 7166 & 1.314 & 2.946 \\ 
        H/h & Secondary TAMS & 3758 & 15.491 & 4057 & 1.175 & 7.834 & 8264 & 1.463 & 2.628 \\
        \bottomrule
    \end{tabular} 
    \label{tab:timestamps}
\end{table*}

\section{Kinematics and Galactic Orbit Parameters} 
\label{sec:kinematics}

The high-precision astrometric data provided by the {\it Gaia} satellite enable the accurate determination of the kinematic properties and Galactic orbital parameters of the solar neighbourhood stars. In this study, the space velocity components and Galactic orbital elements of V570\,Per were calculated using the equatorial coordinates ($\alpha, \delta$), trigonometric parallax ($\varpi$), and proper-motion components ($\mu_{\alpha}\cos\delta$, $\mu_{\delta}$) from the {\it Gaia} DR3 catalogue \citep{Gaia_DR3}, along with the radial velocity ($V_{\gamma}$) of the binary system's center of mass calculated in this study. The astrometric and spectroscopic parameters used in the calculations are summarized in Table~\ref{tab:kinematic}.

To determine the space velocity components of V570\,Per, we utilized the \texttt{galpy} software package developed by \citet{Bovy_2015}. The associated uncertainties were estimated following the procedure described by \citet{Johnson_1987}. The space-velocity components are influenced by the star’s position in the Galaxy as well as observational effects due to the Sun's motion. To minimize such biases, we applied corrections for both Galactic differential rotation and the local standard of rest (LSR). Differential rotation corrections were implemented based on the formulae provided by \citet{Mihalas_1981}, yielding adjustments of $1.85$ km s$^{-1}$ and $-0.24$ km s$^{-1}$ for the $U$ and $V$ velocity components, respectively. Since the $W$ component is not affected by differential rotation, no correction was applied in that case. For the LSR correction, we adopted the solar motion values from \citet{Coskunoglu_2011}, namely $(U, V, W)_{\odot} = (8.83 \pm 0.24, 14.19 \pm 0.34, 6.57 \pm 0.21)$ km s$^{-1}$, and applied them to the differentially corrected space velocities. The total-space velocity relative to the LSR, denoted as $S_{\rm LSR}$, was computed using the relation $S_{\rm LSR} = \sqrt{U_{\rm LSR}^2 + V_{\rm LSR}^2 + W_{\rm LSR}^2}$ \citep{Bilir2012}. The derived velocity values are listed in Table~\ref{tab:kinematic}. To assess the likely Galactic population membership of V570\,Per, we employed the kinematic classification method proposed by \citet{Bensby2003}. The resulting probability distribution indicated that the system has a 99.29\% likelihood of belonging to the thin disk, a 0.70\% likelihood for the thick disk, and only a 0.01\% likelihood of being a halo member. Accordingly, the thick-to-thin disk probability ratio was found to be $TD/D=0.007$ \citep{Bensby2003}. These results strongly suggest that V570\,Per is most likely a member of the Galactic thin-disk population \citep{Leggett_1992, Coskunoglu_2012}.

\begin{table}[t!]
    \setlength{\tabcolsep}{8pt}
    \renewcommand{\arraystretch}{1}
    \small
    \centering
    \caption{Astrometric measurements of V570\,Per together with its radial velocity were used to calculate space velocity components and Galactic orbital parameters. The astrometric and spectroscopic data in the table are taken from the {\it Gaia} DR3 catalogue and this study, respectively. See the first paragraph in \S\ref{sec:kinematics} for an explanation of the input parameters, while the output parameters are discussed in the rest of that section.}
    \begin{tabular}{lr}
        \hline
        \multicolumn{2}{c}{Input Parameters} \\
        \hline
        $\alpha$ (J2000)                            & 03:09:34.94           \\
        $\delta$ (J2000)                            & +48:37:28.69          \\
        $\mu_{\alpha}\cos\delta$ (mas yr$^{-1}$)    & $44.802 \pm 0.034$    \\
        $\mu_{\delta}$ (mas yr$^{-1}$)              & $-39.504 \pm 0.040$   \\
        $\varpi$ (mas)                              & $8.295 \pm 0.036$     \\
        $V_{\rm \gamma}$ (km s$^{-1}$)              & $22.932 \pm 0.123$    \\
        \hline
        \multicolumn{2}{c}{Output Parameters} \\
        \hline
        $U_{\rm LSR}$ (km s$^{-1}$)                 & $-30.39 \pm 0.13$     \\
        $V_{\rm LSR}$ (km s$^{-1}$)                 & $-1.47 \pm 0.14$      \\
        $W_{\rm LSR}$ (km s$^{-1}$)                 & $-3.08 \pm 0.04$      \\
        $S_{\rm LSR}$ (km s$^{-1}$)                 & $30.58 \pm 0.20$      \\
        $R_{\rm a}$ (pc)                            & $8840 \pm 4$          \\
        $R_{\rm p}$ (pc)                            & $7283 \pm 6$          \\
        $Z_{\rm max}$ (pc)                          & $32 \pm 1$            \\
        $e$                                         & $0.095 \pm 0.001$     \\
        $R_{\rm gc}$ (pc)                           & $8098 \pm 2$          \\        
        $R_{\rm birth}$ (pc)                        & $7639 \pm 4$          \\
        \hline
    \end{tabular}
    \label{tab:kinematic}
\end{table}

\begin{figure}[ht!]
\centering\includegraphics[width=1\linewidth]{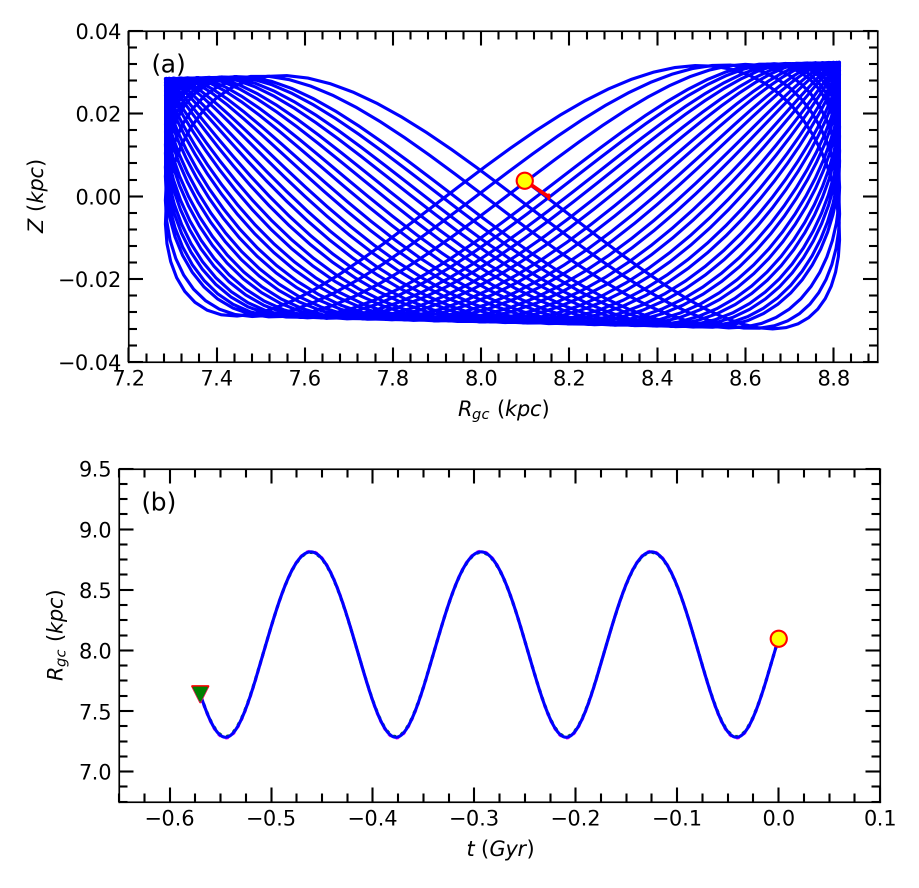}\vspace{-5pt}
\caption{The Galactic orbits and birth radii of V570\,Per in the $Z \times R_{\rm gc}$ (a) and $R_{\rm gc} \times t$ (b) diagrams. The filled yellow circles (today's position) and triangles show the current and birth radius positions of the V570\,Per system in the Galaxy, respectively. The red arrow is the motion vector of V570\,Per today. The green and pink dotted lines show the orbit when errors in input parameters are considered, whereas the green and pink filled triangles represent the birth locations of the V570\,Per based on the lower and upper error estimates.}
\label{fig:galactic_orbits}
\end{figure}

The Galactic orbital parameters of V570\,Per were computed using the \texttt{galpy} code developed by \citet{Bovy_2015}. For these calculations, the Galactic gravitational potential was modelled with \texttt{MWPotential2014}, which is specifically tailored to represent the structure of the Milky Way. To ensure that the system traces a closed orbit around the Galactic center, the orbit was integrated over a timespan of 3 Gyr with time steps of 1.5 Myr. The orbital integration yielded several fundamental parameters, including the apogalactic distance ($R_{\rm a}$), perigalactic distance ($R_{\rm p}$), maximum vertical excursion from the Galactic plane ($Z_{\rm max}$), and orbital eccentricity ($e$). A summary of the derived Galactic orbital parameters is presented in Table~\ref{tab:kinematic}. The system's location concerning the Galactic center ($R_{\rm gc}$) and its vertical distance from the Galactic plane ($Z$) as a function of time are illustrated in Figure~\ref{fig:galactic_orbits}a. Results from the \texttt{galpy} integration indicate that V570\,Per follows a relatively flattened Galactic orbit. In addition, the system's current vertical position of $Z = -17$ pc, computed via the relation $Z = d \times \sin b$, supports its classification as a likely member of the Milky Way's thin-disk population \citep{Guctekin_2019}.

The Galactic orbit of V570\,Per is illustrated in the $Z \times R_{\rm gc}$ and $R_{\rm gc} \times t$ diagrams presented in Figure~\ref{fig:galactic_orbits}. These plots offer an edge-on perspective of the system's trajectory, highlighting its spatial distribution relative to both the Galactic center and the Galactic plane \citep{Tasdemir_2023, Gokmen2023, Tasdemir2025}. In Figure~\ref{fig:galactic_orbits}b, the yellow-filled triangles and circles indicate the estimated birth and current positions of V570\,Per, respectively \citep{Yontan_2022, Akbaba2024, Cinar2024}. The orbit is slightly elliptical, with an eccentricity of 0.09, and the system reaches a maximum vertical distance of $Z_{\rm max} = 32 \pm 1$ pc above the Galactic plane. These kinematic characteristics are consistent with those expected for stars belonging to the old thin-disk population of the Milky Way. To infer the probable birthplace of V570\,Per, its orbit was integrated backward in time over its estimated age of $t=577$ Myr using the \texttt{galpy} framework \citep{Yontan-Canbay_2023, Yucel2024}. This analysis yielded a birth radius of $R_{\rm birth}=7639\pm 4$ pc, suggesting that the system likely originated near the inner, metal-rich region of the Galactic thin disk \citep{Myers2022}.

\section{Discussions} \label{sec:discuss}

In this study, we combined photometric, spectroscopic, and astrometric observations to derive the most precise and detailed parameter estimates to date of the components of the detached binary system V570\,Per. Examining the evolution of systems with nearly equal-mass components, such as this one, is particularly valuable for understanding subtle differences in their evolutionary tracks and the resulting effects on their physical properties \citep{Southworth2013, Alicavus2022, Yucel2022, Alan2024, Alan2025}. For the first time in literature, we have disentangled each spectrum of the components from the composite spectrum using modern techniques, incorporating additional spectra that have not been used in any study of V570\,Per. We have successfully determined chemical abundances for up to 20 elements for each component. Our analysis yields that masses of the components are $1.4569_{-0.0100}^{+0.0094}$ and $1.3579_{-0.0089}^{+0.0094}~M/M_{\odot}$ for the primary and the secondary, radii of the components are $1.543_{-0.009}^{+0.012}$ and $1.377_{-0.015}^{+0.013} ~R/R_{\odot}$, in the same order. The temperatures and the metallicities are $6556_{-26}^{+64}$ K, and $+0.18_{-0.01}^{+0.03}$ dex and $6468_{-21}^{+32}$ K, and $+0.15_{-0.00}^{+0.02}$ dex for the primary and the secondary, respectively. The available O-C data imply a minimum mass of 0.57 $M_{\odot}$ with $P=25974$ days or 0.11 $M_{\odot}$ with $P=6680$ days for the third body, corresponding to eccentric and circular orbits, respectively. Our evolutionary analysis indicates that, assuming they have entered the MS phase at the same time, the initial period of the system was 1.9031 days. The age of the system is currently $577 \pm 60$ Myr, and the orbit is synchronized. The primary component of V570\,Per will fill its Roche lobe in 2.82 Gyr and then start to mass transfer to the secondary component. Our calculations indicate that the secondary will enter the TAMS phase in about 3.2 Gyr. 

The first study in the literature of V570\,Per, in terms of determination of its components, was conducted by \citet{Munari2001} to understand the capabilities of analysis of eclipsing binaries using \textit{Gaia} data. The study concluded that the masses and radii of the primary and secondary components are $1.28\pm0.03$ $M_\odot$ and $1.64\pm0.16$ $R_\odot$, and $1.22\pm0.03$ $M_\odot$ and $1.01\pm0.25$ $R_\odot$, respectively. The temperatures were estimated as $6460\pm150$ K and $6204\pm180$ K, in the same order. Later, \citet{Tomasella2008} used ground-based photometric data in a reanalysis of the system.  They applied constraints for luminosities and used spectroscopic light ratios, based on the existence of shallow eclipses in the light curves. Their analysis yielded component masses of $1.4487\pm0.0055$ $M_\odot$  and $1.3500\pm0.0055$ $M_\odot$, and corresponding radii of $1.523\pm0.030$ $R_\odot$ and $1.388\pm0.019$ $R_\odot$, for the primary and secondary, respectively. By using a grid search, they estimated the temperatures of the components as $6842\pm25$ K and $6562\pm25$ K for the primary component and the secondary component, respectively. They also calculated the metallicity of the system as [M/H]$ = 0.02 \pm 0.03$ dex. \citet{Tomasella2008} compared the parameters obtained for the component stars of V570\,Per using four different evolutionary track computation methods. They reported that the best agreement between the observational data and theoretical models was achieved with the BaSTI \citep{Cordier2007} mass tracks, and estimated the system’s age to be $790\pm60$ Myr. The most current study of V570\,Per was made by \citet{Southworth2023}, using the same RVs of its predecessors and including new photometric data (from \textit{TESS}). Southworth followed the footsteps of \citet{Tomasella2008}: constraining the light ratios and deriving the masses and radii of the components as $1.4489\pm0.0063$ $M_\odot$ and $1.538\pm0.035$ $R_\odot$, and $1.3495\pm0.0062$ $M_\odot$ and $1.349\pm0.032$ $R_\odot$, for the primary and the secondary components, respectively. An analysis was made using the \texttt{PARSEC} evolutionary models \citep{Bressan2012}, leading to the system's age being estimated as being between 800 and 900 Myr. In Table~\ref{tab:comparison}, a comparison of our findings and the literature is given.

Considering the studies in the literature on the age of V570\,Per, it has been found to range between 790 and 900 Myr. However, in the present study, a younger age of $577\pm60$ Myr was derived using \texttt{MESA} evolutionary tracks. One of the main reasons for this discrepancy is the improvement of the codes used in evolutionary track calculations over the past two decades, particularly the transition from single-star to binary-star evolutionary computations. Therefore, the age value obtained in this study can be considered more accurate and reliable. 

\citet{Southworth2023} proposed that the discrepancies between the eclipse timings derived from RV and LC analyses may be attributed to the gravitational influence of a tertiary component within the system. The eclipse timing analysis, supported by both circular and eccentric orbital models (see \S\ref{third_body}). Nevertheless, the currently available data are insufficient to derive the orbital parameters of the putative third body with statistical significance. Continued high-precision timing observations will therefore be essential to constrain its orbital configuration and verify its dynamical influence on the system.

\begin{table*}
	\setlength{\tabcolsep}{0.82pt}
	\renewcommand{\arraystretch}{1.4}
    \small
\centering
\caption{Parameters determined for V570\,Per from the analysis of RV and LC data, and comparison of the results with the literature.} 
\begin{tabular}{lcccccccc}\hline
Parameter                                                      & \multicolumn{2}{c}{\citet{Munari2001}} & \multicolumn{2}{c}{\citet{Tomasella2008}} & \multicolumn{2}{c}{\citet{Southworth2023}} & \multicolumn{2}{c}{This Study}  \\ 
\hline
 & Primary & Secondary & Primary  & Secondary & Primary  & Secondary & Primary  & Secondary \\
\hline
$a$ ($R_\odot$)                     & \multicolumn{2}{c}{$8.75^{+0.06}_{-0.06}$} & \multicolumn{2}{c}{$9.0929^{+0.0120}_{-0.0120}$} & \multicolumn{2}{c}{$9.100^{+0.013}_{-0.013}$} & \multicolumn{2}{c}{$9.118^{+0.016}_{-0.015}$}    \\
 \emph{q}                           & \multicolumn{2}{c}{$0.951^{+0.012}_{-0.012}$} & \multicolumn{2}{c}{$0.9319^{+0.0008}_{-0.0008}$} & \multicolumn{2}{c}{$0.9314^{+0.0026}_{-0.0026}$} & \multicolumn{2}{c}{$0.932^{+0.003}_{-0.003}$}  \\
\(V_\gamma\)   (km s$^{-1}$)        & \multicolumn{2}{c}{$21.91^{+0.45}_{-0.45}$} & \multicolumn{2}{c}{$22.81^{+0.03}_{-0.03}$} & \multicolumn{2}{c}{--}       & \multicolumn{2}{c}{$22.932^{+0.126}_{-0.120}$}  \\
\emph{e}                            & \multicolumn{2}{c}{$0$} & \multicolumn{2}{c}{$0$} & \multicolumn{2}{c}{$0$ (fixed)} & \multicolumn{2}{c}{0 (fixed)} \\
\emph{i}  ($^\circ$)                            & \multicolumn{2}{c}{$78.8^{+1.02}_{-1.02}$} & \multicolumn{2}{c}{$77.44^{+0.32}_{-0.32}$} & \multicolumn{2}{c}{$77.294^{+0.048}_{-0.048}$} & \multicolumn{2}{c}{$77.170^{+0.072}_{-0.073}$} \\
$T_\mathrm{eff,b}/T_\mathrm{eff,a}$ & \multicolumn{2}{c}{$0.960^{+0.051}_{-0.049}$} & \multicolumn{2}{c}{$0.959^{+0.009}_{-0.009}$} & \multicolumn{2}{c}{$0.959^{+0.014}_{-0.014}$} & \multicolumn{2}{c}{$0.962^{+0.003}_{-0.003}$}   \\
\emph{M} ($M_\odot$)                & $1.28^{+0.03}_{-0.03}$ & $1.22^{+0.03}_{-0.03}$ & $1.4487^{+0.0058}_{-0.0058}$   & $1.3500^{+0.0055}_{-0.0055}$ & $1.4489^{+0.0063}_{-0.0063}$   & $1.3495^{+0.0062}_{-0.0062}$ & $1.4569^{+0.0100}_{-0.0094}$ & $1.3579^{+0.0094}_{-0.0089}$ \\
\emph{R} ($R_\odot$)                & $1.64^{+0.16}_{-0.16}$      & $1.01^{+0.25}_{-0.25}$ & $1.523^{+0.030}_{-0.030}$      & $1.388^{+0.019}_{-0.019}$  & $1.538^{+0.035}_{-0.035}$      & $1.349^{+0.032}_{-0.032}$ & $1.543^{+0.012}_{-0.009}$      & $1.377^{+0.013}_{-0.015}$    \\
$\log g$ (cgs)                      & $4.12^{+0.05}_{-0.05}$  & $4.55^{+0.12}_{-0.12}$ & $4.23^{+0.02}_{-0.02}$    & $4.28^{+0.01}_{-0.01}$ & $4.225^{+0.020}_{-0.020}$    & $4.308^{+0.021}_{-0.021}$ & $4.225^{+0.009}_{-0.009}$      & $4.293^{+0.013}_{-0.011}$   \\
\hline
\label{tab:comparison}
\end{tabular}
\end{table*}

\begin{figure}
    \centering
    \includegraphics[width=1\linewidth]{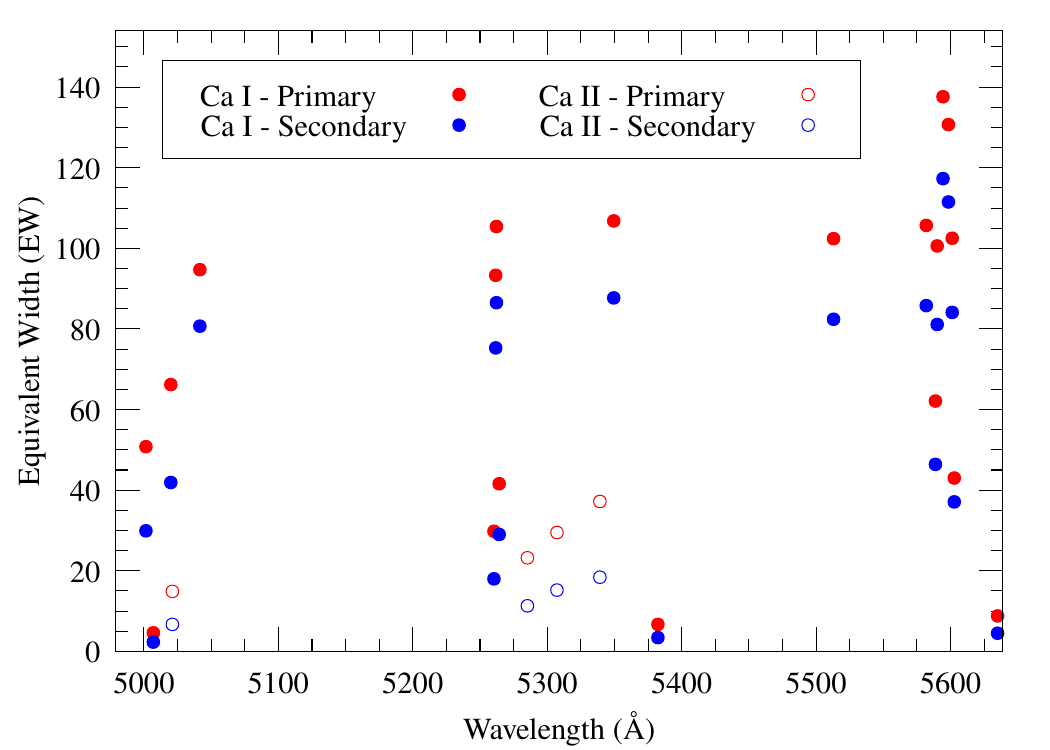}
    \caption{Measured EW of Ca for the primary and secondary components by \texttt{SP\_Ace}. Neutral Ca (Ca-I) and ionized Ca (Ca-II) were represented with filled and empty circles, respectively.}
    \label{fig:ca_ew}
    \end{figure}

One of the main goals of this study is to determine the temperatures and chemical abundances of the components of V570\,Per independently. To this end, we disentangled the composite spectrum using a total of 34 spectra and derived the chemical abundances of each component with \texttt{SP\_Ace}. As shown in Table~\ref{tab:chem_results} and Figure~\ref{fig:abundances}, the abundances of all elements are consistent within the uncertainties, as expected since both components formed in the same molecular cloud. The most significant deviation is observed in calcium (Ca). \texttt{SP\_Ace} treats Ca as a high-quality element \citep[see p.12 in][]{space2}, which indicates that its measurements for Ca should be reliable. To verify this, we examined the equivalent widths (EWs) of Ca lines (Figure~\ref{fig:ca_ew}) to check for possible systematic trends. The results clearly show that all Ca EWs in the primary component are larger than those in the secondary component, indicating a higher Ca abundance in the primary star. If similar discrepancies were present in other elements, it could suggest an issue with the disentangling procedure; however, no such trend is observed. \texttt{SP\_Ace} calculates chemical abundances based on the assumption of local thermodynamic equilibrium (LTE). The calculation of chemical abundance of Ca based on LTE or a non-LTE assumption may yield differences of up to 0.1 dex \citep{Osorio2020, Lind2024}. Our result has a difference of more than 0.3 dex, which cannot be explained by such a difference in assumptions. Therefore, we conclude that Ca is genuinely more abundant in the primary component than in the secondary component.

\begin{figure*}[ht]
\centering\includegraphics[width=1\linewidth]{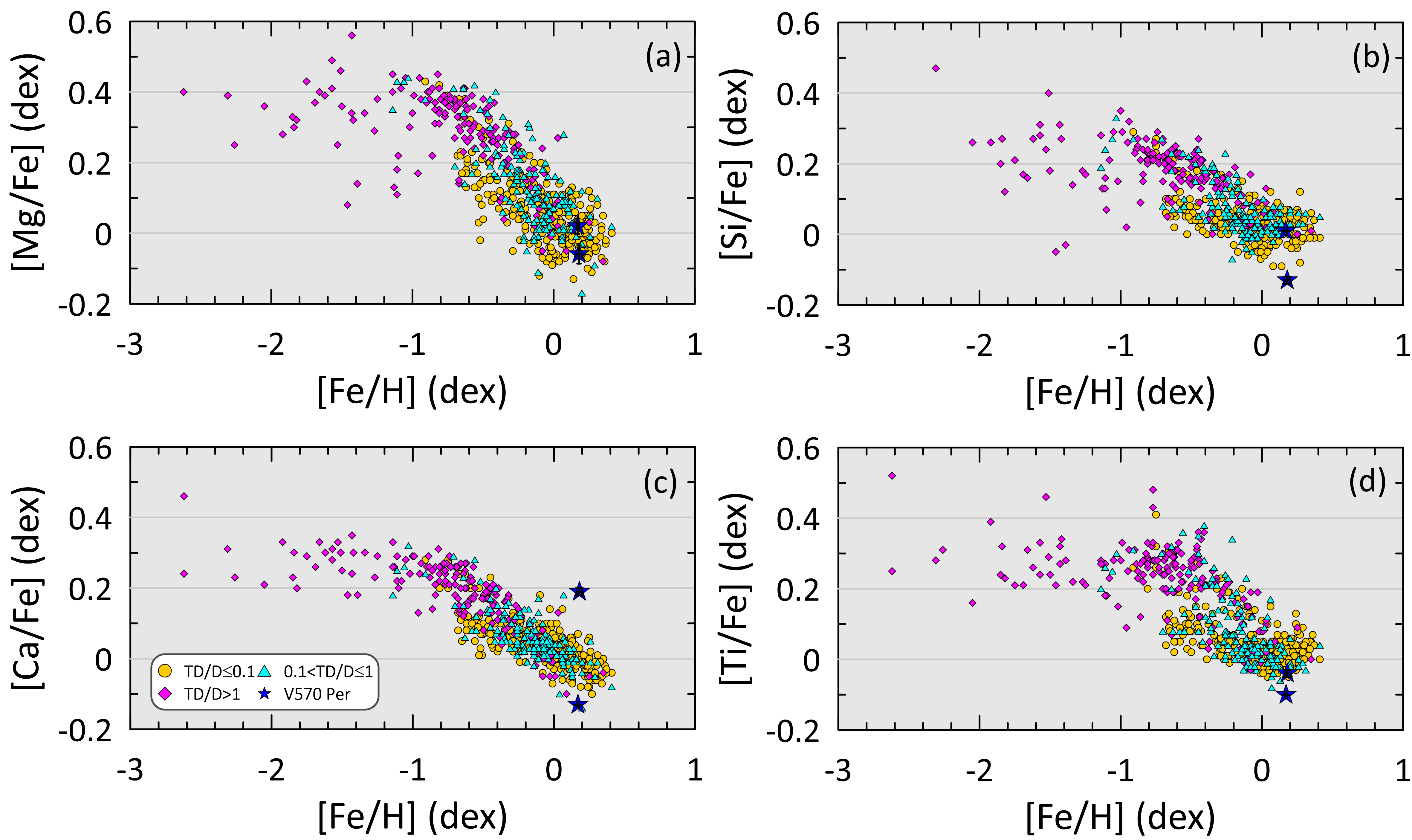}\vspace{-5pt}
\caption{The abundances of four $\alpha$-elements relative to iron are presented. The panels compare the component stars of V570\,Per (blue star symbols) with the stellar populations (thin disk, thick disk, and halo) from \citet{Bensby2014}, which were classified based on kinematic properties.}
\label{fig:Alpha_elements}
\end{figure*} 

In determining the Galactic population membership of V570\,Per, both the elemental abundances of $\alpha$ and iron-peak elements as well as the kinematic properties of its components were considered. Since no detailed abundance analysis of V570\,Per has previously been reported in the literature, the abundances derived for its three components together with the system’s kinematic parameters were evaluated in comparison with the 714 F- and G-type stars from the solar neighborhood analyzed by \citet{Bensby2014}. In this context, the abundances of four $\alpha$-elements such as Mg, Si, Ca, and Ti measured in the spectra of both components of V570\,Per were plotted in the [X/Fe]-[Fe/H] chemical planes (Figure~\ref{fig:Alpha_elements}), alongside the reference sample of \citet{Bensby2014}. The latter study classifies stars into three populations based on kinematic criteria: high-probability thin disk stars ($TD/D \leq 0.1$), low-probability thin disk stars ($0.1 < TD/D \leq 1$), and stars with thick-disk or halo kinematics ($TD/D > 1$). Except for the relatively high-calcium abundance measured in the primary component of V570\,Per (see Table~\ref{tab:chem_results}), the abundances of the other $\alpha$-elements are in good agreement with the chemical and kinematic distributions reported by \citet{Bensby2014}. The generally low $\alpha$-element abundances, combined with [Fe/H] values exceeding the solar metallicity, suggest that V570\,Per is chemically consistent with the thin-disk population. Moreover, the system’s kinematic parameters align with stars exhibiting $TD/D < 1$, further reinforcing this classification. Finally, the deficiency of $\alpha$-elements and the relative enrichment of iron in the component stars imply that the molecular cloud from which V570\,Per formed was predominantly enriched by Type Ia supernovae \citep[e.g.,][]{Gratton2000, Matteucci2009, Maoz2017}.

\begin{figure}[ht]
\centering\includegraphics[width=1\linewidth]{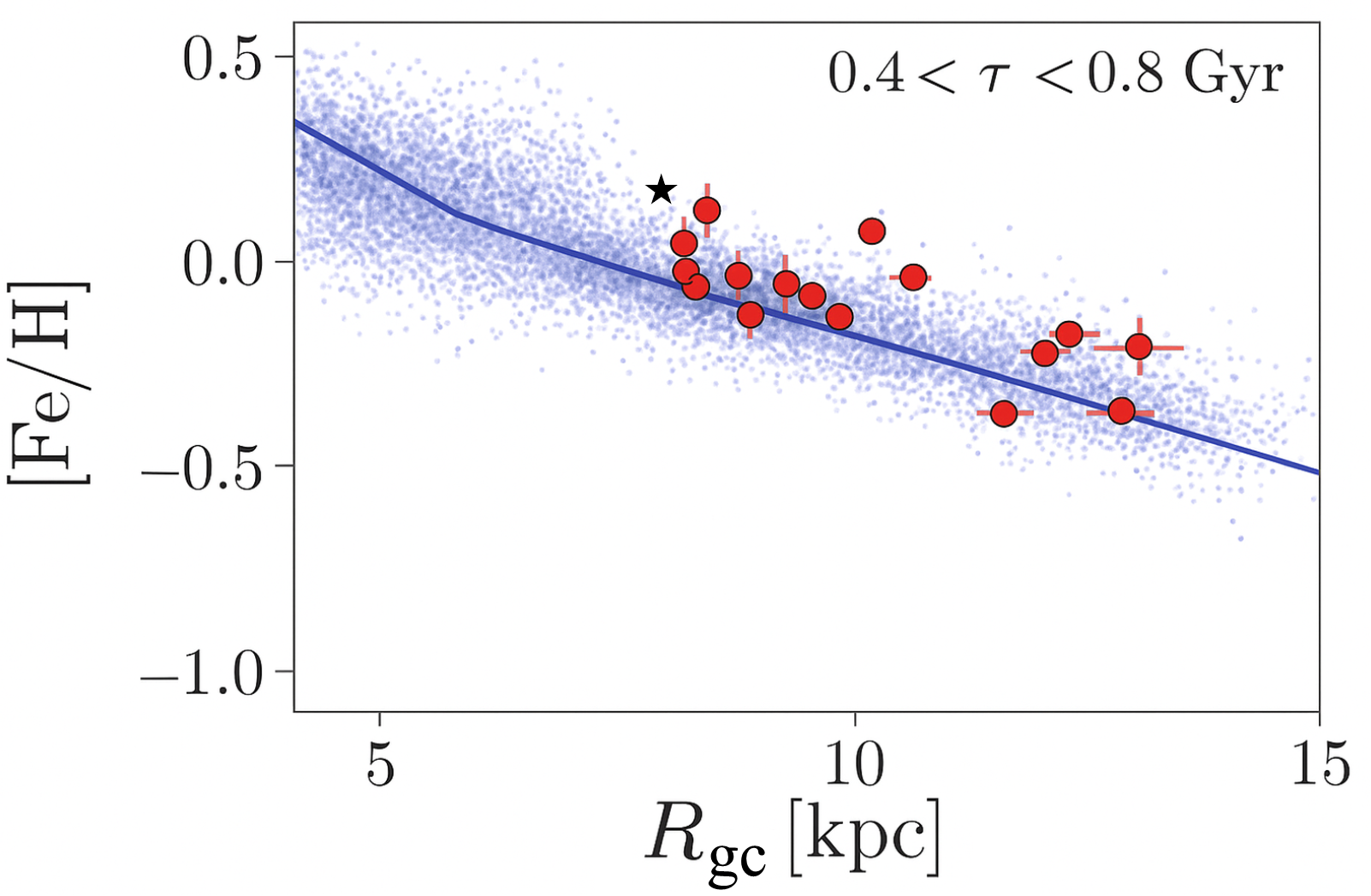}\vspace{-5pt}
\caption{V570\,Per in comparison with the chemical evolution model of \citet{Chiappini2009} (blue) and the chemo-dynamical simulations by \citet{Minchev2013, Minchev2014} (blue dot). Red circles and black star symbol are represented by open clusters taken from \citet{Myers2022} and V570\,Per, respectively.}
\label{fig:FeH_Rgc}
\end{figure}

Using the mean metallicity derived for the V570\,Per system ($\langle{\rm [Fe/H]}\rangle = 0.175$ dex) and its Galactocentric distance ($R_{\rm gc}\cong 8.1$ kpc), the system was compared with stars of similar ages in the solar neighborhood. For this comparison, we employed the thin-disk chemical evolution model developed by \citet{Chiappini2009} as well as the chemo-dynamical simulations of \citet{Minchev2013, Minchev2014}. In addition, the analysis incorporated data from 16 open clusters presented by \citet{Myers2022}, based on spectroscopic studies of cluster member stars in the APOGEE DR17 catalog \citep{Abdurrouf2022}. The distribution of metallicity versus Galactocentric distance for V570\,Per is shown in Figure~\ref{fig:FeH_Rgc}. As seen in the figure, V570\,Per is located in a region more metal-rich than the simulated field stars. This result is consistent with the findings for the 16 open clusters studied by \citet{Myers2022}. Consequently, it suggests that V570\,Per may have originated in a metal-rich, young-open cluster and subsequently dispersed into the field star population.

To further investigate this possibility, we examined the sky region surrounding V570\,Per. A literature search using the SIMBAD database indicates that V570\,Per has been classified as a member of the open cluster Melotte 20 (Collinder 39, $\alpha$ Per; $l = 147^{\rm o}.4710, b=-6^{\rm o}.4620$) by \citet{Heckmann1956}. Astrometric analyses of Melotte 20 based on {\it Gaia} DR3 data report proper motion components ($\mu_{\alpha}\cos \delta=22.9182$, $\mu_{\delta}=-25.4626$ mas yr$^{-1}$), a distance of $175.58^{+0.04}_{-0.03}$ pc, an age of $56^{+68}_{-26}$ Myr, a radial velocity of $-0.87 \pm 1.25$ km s$^{-1}$ \citep{Hunt2023} and a mean metallicity of $0.158 \pm 0.070$ dex \citep{Dias2021}. When the astrometric parameters and fundamental parameters determined for V570\,Per in this study (see Table~\ref{table:fund}) are compared with the values compiled from the literature for Melotte 20, a significant discrepancy is found except for the metallicity value. This indicates that V570\,Per is not a former member of Melotte 20. Therefore, V570\,Per may have originated in a different open cluster, or it may have formed within a metal-rich molecular cloud independent of any cluster environment.

\begin{figure}[ht]
    \centering
    \includegraphics[width=0.9\linewidth]{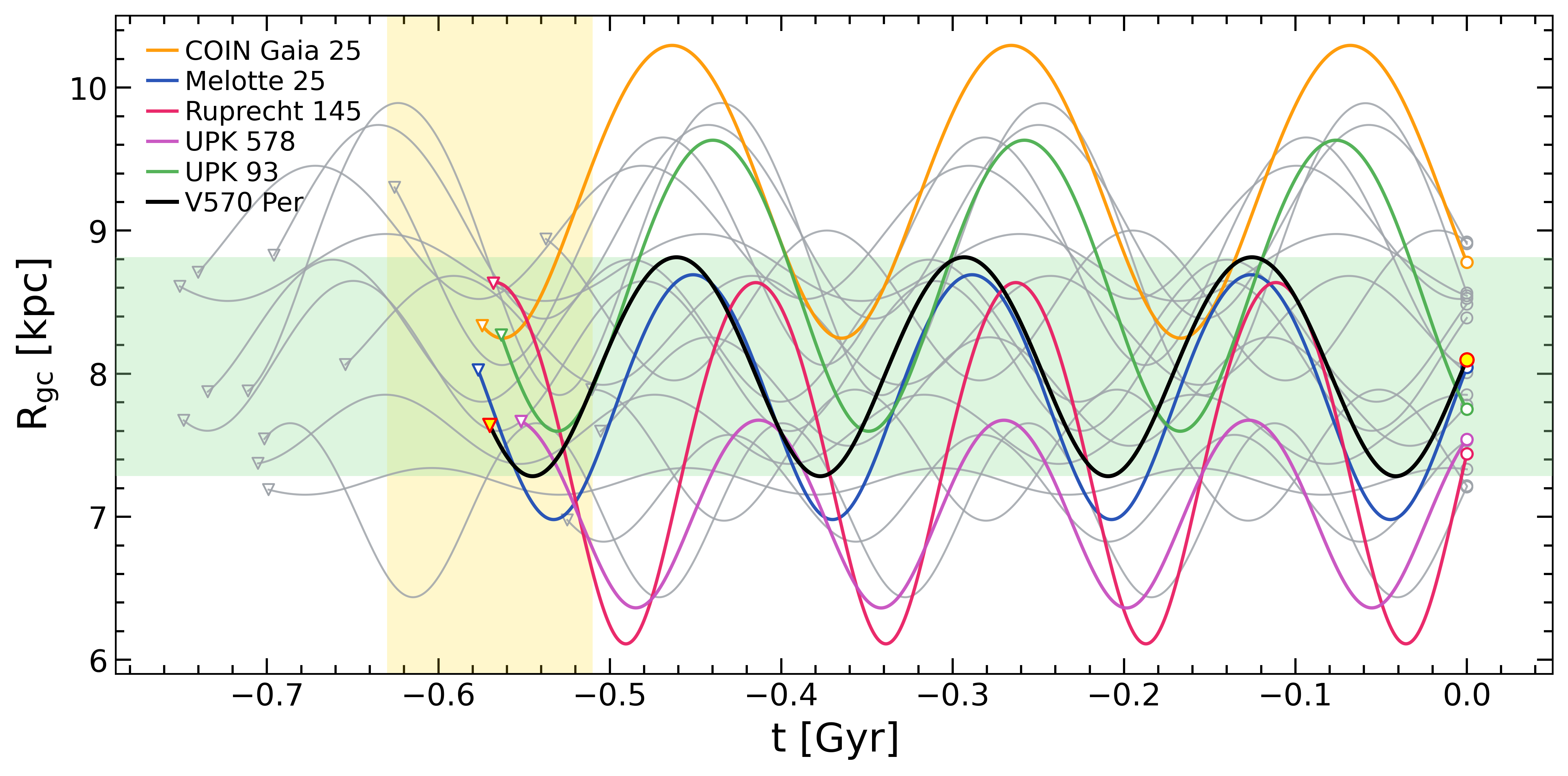}\vspace{-5pt}
    \caption{Possible birthplace calculated by tracing back in time the positions of 20 open clusters selected from the \citet{Hunt2023} catalog and the V570\,Per system, considering their ages. The yellow and green shaded areas represent, respectively, the age of the V570\,Per system, and the corresponding possible birthplace. The open clusters shown in five different colors indicate the potential birthplace of the V570\,Per system.}
    \label{fig:Rgc_time_birth}
\end{figure}

To determine in which open cluster V570\,Per was formed in the solar neighborhood, this study aimed to identify open clusters whose ages and possible birthplace correspond to the age estimated for the system ($t = 577\pm 60$ Myr) and its associated range of birth radii ($7.3 < R_{\rm gc} ({\rm kpc}) < 8.9$) (see Figure~\ref{fig:galactic_orbits}). Candidate open clusters were selected from the catalog of \citet{Hunt2023}, which provides analyses based on {\it Gaia} DR3 data. From this catalog, 20 open clusters located within 1 kpc of the Sun and with ages in the range of 500–800 Myr were chosen. The possible birthplaces ($R_{\rm gc}$) of the selected open clusters were computed using the method described in \S\ref{sec:kinematics} and \citet{Akbaba2024} study, depending on the open cluster ages. This approach simultaneously accounts for both spatial proximity and evolutionary consistency, allowing the identification of the most probable open cluster associated with the V570\,Per system. To construct the Galactic orbits of the selected open clusters, their astrometric and spectroscopic data were taken from the catalog of \citet{Hunt2023}. By tracing back the motions of the selected clusters and V570\,Per according to their ages, the corresponding birth radii of the objects were determined, and their Galactic orbital motions are illustrated in Figure~\ref{fig:Rgc_time_birth}. Considering the uncertainties in the age of V570\,Per and the corresponding variations in $R_{\rm gc}$, possible candidate open clusters in which the system may have formed were identified.

Figure~\ref{fig:Rgc_time_birth} shows five open clusters whose orbits intersect the overlap region of the two shaded areas, indicating that they are dynamically consistent with V570\,Per. The ages of these clusters, adopted from the catalog of \citet{Hunt2023}, and the corresponding $R_{\rm gc}$ values derived from the dynamical orbit analysis are as follows: COIN-Gaia\,25 ($575^{+353}_{-207}$ Myr, 8.34 kpc), Melotte\,25 ($577^{+415}_{-221}$ Myr, 8.04 kpc), Ruprecht\,145 ($568^{+277}_{-216}$ Myr, 8.63 kpc), UPK\,578 ($552^{+258}_{-234}$ Myr, 7.68 kpc), and UPK\,93 ($563^{+378}_{-211}$ Myr, 8.28 kpc). An additional parameter used to determine the possible birth cluster of the V570\,Per system was the mean metallicity of the candidate clusters, derived from spectroscopic analyses available in the literature. Accordingly, COIN-Gaia\,25, Melotte\,25, and Ruprecht\,145 were analyzed by \citet{Spina2021} using GALAH data; UPK\,578 was studied by \citet{Andrea2023} using {\it Gaia} XP spectra data; and UPK\,93 was analyzed by \citet{Fu2022} based on LAMOST observations. According to these spectroscopic analyses, the mean metallicities were determined as follows: COIN-Gaia\,25 (from 1 member star) $\langle[\rm Fe/H]\rangle=0.007\pm 0.008$ dex, Melotte\,25 (from 34 member stars) $\langle[\rm Fe/H]\rangle=0.145\pm 0.063$ dex, Ruprecht\,145 (from 2  member stars) $\langle[\rm Fe/H]\rangle=0.007\pm 0.008$ dex, UPK\,578 (from 57 member stars) $\langle[\rm Fe/H]\rangle=-0.080\pm 0.080$ dex, and UPK\,93 (from 2 member stars) $\langle[\rm Fe/H]\rangle=-0.007\pm 0.005$ dex. Considering the spectroscopic properties of the five candidate open clusters, the metallicity of V570\,Per determined in this study ([Fe/H] = 0.175$^{+0.020}_{-0.010}$ dex) shows a strong agreement with that of Melotte\,25. This finding provides significant evidence that V570\,Per may have originated in the Melotte\,25 open cluster.

\section*{Acknowledgments}

We thank the anonymous referee for their insightful and constructive suggestions that significantly improved the paper. We thank T\"{U}B\.{I}TAK for funding this research under project number: 123C161. Funding was provided by the Scientific Research Projects Coordination Unit of Istanbul University as project number 40044. We made use of spectral data retrieved from the ELODIE archive at Observatoire de Haute-Provence (OHP). Some of our spectral data are based on observations collected at Copernico 1.82m telescope (Asiago Mount Ekar, Italy) INAF -- Osservatorio Astronomico di Padova. We are deeply thankful for all the efforts of Dr.\  Lina Tomasella to provide us with the ASIAGO data. This research has made use of the Astrophysics Data System, funded by NASA under Cooperative Agreement 80NSSC25M7105. The VizieR and Simbad databases at CDS, Strasbourg, France were invaluable for the project as were data from the European Space Agency (ESA) mission \emph{Gaia}\footnote{https://www.cosmos.esa.int/gaia}, processed by the \emph{Gaia} Data Processing and Analysis Consortium (DPAC)\footnote{https://www.cosmos.esa.int/web/gaia/dpac/consortium}. Funding for DPAC has been provided by national institutions, in particular, the institutions participating in the \emph{Gaia} Multilateral Agreement. This paper includes data collected with the TESS mission, obtained from the MAST data archive at the Space Telescope Science Institute (STScI). Funding for the TESS mission is provided by the NASA Explorer Program. STScI is operated by the Association of Universities for Research in Astronomy, Inc., under NASA contract NAS 5–26555. The data described here may be obtained from the MAST archive at \dataset[doi:10.17909/fwdt-2x66]{https://dx.doi.org/10.17909/fwdt-2x66}. This research made use of Lightkurve, a Python package for Kepler and TESS data analysis \citep{lk}.

\software{
\texttt{Astropy} \citep{astro1,astro2,astro3},
\texttt{Astroquery} \citep{Astroquery},
\texttt{ATLAS9} \citep{Kurucz1979,Castelli2003}, 
\texttt{corner} \citep{corner}, 
\texttt{galpy} \citep{Bovy_2015}, 
\texttt{IRAF} \citep{Tody1986, Tody1993},
\texttt{lightkurve} \citep{lk},
\texttt{Matplotlib} \citep{matplotlib}, 
\texttt{MESA} \citep{Paxton2011, Paxton2013, Paxton2015, Paxton2018, Paxton2019, Jermyn2023},
\texttt{MESA SDK} \citep{SDK},
\texttt{MWPotential2014} \citep{Bovy_2015},
\texttt{NumPy} \citep{numpy}, 
\texttt{PHOEBE} \citep{phoebe1}, 
\texttt{SciPy} \citep{scipy},
\texttt{SP\_Ace} \citep{space1,space2}, and
\texttt{SPECTRUM} \citep{Gray1994}.
} 

\bibliography{reference}
\bibliographystyle{aasjournalv7}



\end{document}